June 2009

# The Cosmic Ray Energy Spectrum and Related Measurements with the Pierre Auger Observatory

Presentations for the
31st International Cosmic Ray Conference, Łódź , Poland, July 2009

# PIERRE AUGER COLLABORATION

J. Abraham[8], P. Abreu[71], M. Aglietta[54], C. Aguirre[12], E.J. Ahn[87], D. Allard[31], I. Allekotte[1], J. Allen[90], J. Alvarez-Muñiz[78], M. Ambrosio[48], L. Anchordoqui[104], S. Andringa[71], A. Anzalone[53], C. Aramo[48], E. Arganda[75], S. Argirò[51], K. Arisaka[95], F. Arneodo[55], F. Arqueros[75], T. Asch[38], H. Asorey[1], P. Assis[71], J. Aublin[33], M. Ave[96], G. Avila[10], T. Bäcker[42], D. Badagnani[6], K.B. Barber[11], A.F. Barbosa[14], S.L.C. Barroso[20], B. Baughman[92], P. Bauleo[85], J.J. Beatty[92], T. Beau[31], B.R. Becker[101], K.H. Becker[36], A. Bellétoile[34], J.A. Bellido[11, 93], S. BenZvi[103], C. Berat[34], P. Bernardini[47], X. Bertou[1], P.L. Biermann[39], P. Billoir[33], O. Blanch-Bigas[33], F. Blanco[75], C. Bleve[47], H. Blümer[41, 37], M. Boháčová[96, 27], D. Boncioli[49], C. Bonifazi[33], R. Bonino[54], N. Borodai[69], J. Brack[85], P. Brogueira[71], W.C. Brown[86], R. Bruijn[81], P. Buchholz[42], A. Bueno[77], R.E. Burton[83], N.G. Busca[31], K.S. Caballero-Mora[41], L. Caramete[39], R. Caruso[50], W. Carvalho[17], A. Castellina[54], O. Catalano[53], L. Cazon[96], R. Cester[51], J. Chauvin[34], A. Chiavassa[54], J.A. Chinellato[18], A. Chou[87, 90], J. Chudoba[27], J. Chye[89d], R.W. Clay[11], E. Colombo[2], R. Conceição[71], B. Connolly[102], F. Contreras[9], J. Coppens[65, 67], A. Cordier[32], U. Cotti[63], S. Coutu[93], C.E. Covault[83], A. Creusot[73], A. Criss[93], J. Cronin[96], A. Curutiu[39], S. Dagoret-Campagne[32], R. Dallier[35], K. Daumiller[37], B.R. Dawson[11], R.M. de Almeida[18], M. De Domenico[50], C. De Donato[46], S.J. de Jong[65], G. De La Vega[8], W.J.M. de Mello Junior[18], J.R.T. de Mello Neto[23], I. De Mitri[47], V. de Souza[16], K.D. de Vries[66], G. Decerprit[31], L. del Peral[76], O. Deligny[30], A. Della Selva[48], C. Delle Fratte[49], H. Dembinski[40], C. Di Giulio[49], J.C. Diaz[89], P.N. Diep[105], C. Dobrigkeit [18], J.C. D'Olivo[64], P.N. Dong[105], A. Dorofeev[88], J.C. dos Anjos[14], M.T. Dova[6], D. D'Urso[48], I. Dutan[39], M.A. DuVernois[98], R. Engel[37], M. Erdmann[40], C.O. Escobar[18], A. Etchegoyen[2], P. Facal San Luis[96, 78], H. Falcke[65, 68], G. Farrar[90], A.C. Fauth[18], N. Fazzini[87], F. Ferrer[83], A. Ferrero[2], B. Fick[89], A. Filevich[2], A. Filipčič[72, 73], I. Fleck[42], S. Fliescher[40], C.E. Fracchiolla[85], E.D. Fraenkel[66], W. Fulgione[54], R.F. Gamarra[2], S. Gambetta[44], B. García[8], D. García Gámez[77], D. Garcia-Pinto[75], X. Garrido[37, 32], G. Gelmini[95], H. Gemmeke[38], P.L. Ghia[30, 54], U. Giaccari[47], M. Giller[70], H. Glass[87], L.M. Goggin[104], M.S. Gold[101], G. Golup[1], F. Gomez Albarracin[6], M. Gómez Berisso[1], P. Gonçalves[71], M. Gonçalves do Amaral[24], D. Gonzalez[41], J.G. Gonzalez[77, 88], D. Góra[41, 69], A. Gorgi[54], P. Gouffon[17], S.R. Gozzini[81], E. Grashorn[92], S. Grebe[65], M. Grigat[40], A.F. Grillo[55], Y. Guardincerri[4], F. Guarino[48], G.P. Guedes[19], J. Gutiérrez[76], J.D. Hague[101], V. Halenka[28], P. Hansen[6], D. Harari[1], S. Harmsma[66, 67], J.L. Harton[85], A. Haungs[37], M.D. Healy[95], T. Hebbeker[40], G. Hebrero[76], D. Heck[37], V.C. Holmes[11], P. Homola[69], J.R. Hörandel[65], A. Horneffer[65], M. Hrabovský[28, 27], T. Huege[37], M. Hussain[73], M. Iarlori[45], A. Insolia[50], F. Ionita[96], A. Italiano[50], S. Jiraskova[65], M. Kaducak[87], K.H. Kampert[36], T. Karova[27], P. Kasper[87], B. Kégl[32], B. Keilhauer[37], E. Kemp[18], R.M. Kieckhafer[89], H.O. Klages[37], M. Kleifges[38], J. Kleinfeller[37], R. Knapik[85], J. Knapp[81], D.-H. Koang[34], A. Krieger[2], O. Krömer[38], D. Kruppke-Hansen[36], F. Kuehn[87], D. Kuempel[36], N. Kunka[38], A. Kusenko[95], G. La Rosa[53], C. Lachaud[31], B.L. Lago[23], P. Lautridou[35], M.S.A.B. Leão[22], D. Lebrun[34], P. Lebrun[87], J. Lee[95], M.A. Leigui de Oliveira[22], A. Lemiere[30], A. Letessier-Selvon[33], M. Leuthold[40], I. Lhenry-Yvon[30], R. López[59], A. Lopez Agüera[78], K. Louedec[32], J. Lozano Bahilo[77], A. Lucero[54], H. Lyberis[30], M.C. Maccarone[53], C. Macolino[45], S. Maldera[54], D. Mandat[27], P. Mantsch[87], A.G. Mariazzi[6], I.C. Maris[41], H.R. Marquez Falcon[63], D. Martello[47], O. Martínez Bravo[59], H.J. Mathes[37], J. Matthews[88, 94], J.A.J. Matthews[101], G. Matthiae[49], D. Maurizio[51], P.O. Mazur[87], M. McEwen[76], R.R. McNeil[88], G. Medina-Tanco[64], M. Melissas[41], D. Melo[51], E. Menichetti[51], A. Menshikov[38], R. Meyhandan[14], M.I. Micheletti[2], G. Miele[48], W. Miller[101], L. Miramonti[46], S. Mollerach[1], M. Monasor[75], D. Monnier Ragaigne[32], F. Montanet[34], B. Morales[64], C. Morello[54], J.C. Moreno[6], C. Morris[92], M. Mostafá[85], C.A. Moura[48], S. Mueller[37], M.A. Muller[18], R. Mussa[51], G. Navarra[54], J.L. Navarro[77], S. Navas[77], P. Necesal[27], L. Nellen[64], C. Newman-Holmes[87], D. Newton[81], P.T. Nhung[105], N. Nierstenhoefer[36], D. Nitz[89], D. Nosek[26], L. Nožka[27], M. Nyklicek[27], J. Oehlschläger[37], A. Olinto[96], P. Oliva[36], V.M. Olmos-Gilbaja[78], M. Ortiz[75], N. Pacheco[76], D. Pakk Selmi-Dei[18], M. Palatka[27], J. Pallotta[3], G. Parente[78], E. Parizot[31], S. Parlati[55], S. Pastor[74], M. Patel[81], T. Paul[91], V. Pavlidou[96c], K. Payet[34], M. Pech[27], J. Pękala[69], I.M. Pepe[21], L. Perrone[52], R. Pesce[44], E. Petermann[100], S. Petrera[45], P. Petrinca[49], A. Petrolini[44], Y. Petrov[85], J. Petrovic[67], C. Pfendner[103], R. Piegaia[4], T. Pierog[37], M. Pimenta[71], T. Pinto[74], V. Pirronello[50], O. Pisanti[48], M. Platino[2], J. Pochon[1], V.H. Ponce[1], M. Pontz[42], P. Privitera[96], M. Prouza[27], E.J. Quel[3], J. Rautenberg[36], O. Ravel[35], D. Ravignani[2],

A. Redondo[76], B. Revenu[35], F.A.S. Rezende[14], J. Ridky[27], S. Riggi[50], M. Risse[36], C. Rivière[34], V. Rizi[45], C. Robledo[59], G. Rodriguez[49], J. Rodriguez Martino[50], J. Rodriguez Rojo[9], I. Rodriguez-Cabo[78], M.D. Rodríguez-Frías[76], G. Ros[75], [76], J. Rosado[75], T. Rossler[28], M. Roth[37], B. Rouillé-d'Orfeuil[31], E. Roulet[1], A.C. Rovero[7], F. Salamida[45], H. Salazar[59b], G. Salina[49], F. Sánchez[64], M. Santander[9], C.E. Santo[71], E.M. Santos[23], F. Sarazin[84], S. Sarkar[79], R. Sato[9], N. Scharf[40], V. Scherini[36], H. Schieler[37], P. Schiffer[40], A. Schmidt[38], F. Schmidt[96], T. Schmidt[41], O. Scholten[66], H. Schoorlemmer[65], J. Schovancova[27], P. Schovánek[27], F. Schroeder[37], S. Schulte[40], F. Schüssler[37], D. Schuster[84], S.J. Sciutto[6], M. Scuderi[50], A. Segreto[53], D. Semikoz[31], M. Settimo[47], R.C. Shellard[14], [15], I. Sidelnik[2], B.B. Siffert[23], A. Śmiałkowski[70], R. Šmída[27], B.E. Smith[81], G.R. Snow[100], P. Sommers[93], J. Sorokin[11], H. Spinka[82], [87], R. Squartini[9], E. Strazzeri[32], A. Stutz[34], F. Suarez[2], T. Suomijärvi[30], A.D. Supanitsky[64], M.S. Sutherland[92], J. Swain[91], Z. Szadkowski[70], A. Tamashiro[7], A. Tamburro[41], T. Tarutina[6], O. Taşcău[36], R. Tcaciuc[42], D. Tcherniakhovski[38], D. Tegolo[58], N.T. Thao[105], D. Thomas[85], R. Ticona[13], J. Tiffenberg[4], C. Timmermans[67], [65], W. Tkaczyk[70], C.J. Todero Peixoto[22], B. Tomé[71], A. Tonachini[51], I. Torres[59], P. Travnicek[27], D.B. Tridapalli[17], G. Tristram[31], E. Trovato[50], M. Tueros[6], R. Ulrich[37], M. Unger[37], M. Urban[32], J.F. Valdés Galicia[64], I. Valiño[37], L. Valore[48], A.M. van den Berg[66], J.R. Vázquez[75], R.A. Vázquez[78], D. Veberič[73], [72], A. Velarde[13], T. Venters[96], V. Verzi[49], M. Videla[8], L. Villaseñor[63], S. Vorobiov[73], L. Voyvodic[87‡], H. Wahlberg[6], P. Wahrlich[11], O. Wainberg[2], D. Warner[85], A.A. Watson[81], S. Westerhoff[103], B.J. Whelan[11], G. Wieczorek[70], L. Wiencke[84], B. Wilczyńska[69], H. Wilczyński[69], C. Wileman[81], M.G. Winnick[11], H. Wu[32], B. Wundheiler[2], T. Yamamoto[96a], P. Younk[85], G. Yuan[88], A. Yushkov[48], E. Zas[78], D. Zavrtanik[73], [72], M. Zavrtanik[72], [73], I. Zaw[90], A. Zepeda[60b], M. Ziolkowski[42]

*[1] Centro Atómico Bariloche and Instituto Balseiro (CNEA-UNCuyo-CONICET), San Carlos de Bariloche, Argentina*

*[2] Centro Atómico Constituyentes (Comisión Nacional de Energía Atómica/CONICET/UTN- FRBA), Buenos Aires, Argentina*

*[3] Centro de Investigaciones en Láseres y Aplicaciones, CITEFA and CONICET, Argentina*

*[4] Departamento de Física, FCEyN, Universidad de Buenos Aires y CONICET, Argentina*

*[6] IFLP, Universidad Nacional de La Plata and CONICET, La Plata, Argentina*

*[7] Instituto de Astronomía y Física del Espacio (CONICET), Buenos Aires, Argentina*

*[8] National Technological University, Faculty Mendoza (CONICET/CNEA), Mendoza, Argentina*

*[9] Pierre Auger Southern Observatory, Malargüe, Argentina*

*[10] Pierre Auger Southern Observatory and Comisión Nacional de Energía Atómica, Malargüe, Argentina*

*[11] University of Adelaide, Adelaide, S.A., Australia*

*[12] Universidad Catolica de Bolivia, La Paz, Bolivia*

*[13] Universidad Mayor de San Andrés, Bolivia*

*[14] Centro Brasileiro de Pesquisas Fisicas, Rio de Janeiro, RJ, Brazil*

*[15] Pontifícia Universidade Católica, Rio de Janeiro, RJ, Brazil*

*[16] Universidade de São Paulo, Instituto de Física, São Carlos, SP, Brazil*

*[17] Universidade de São Paulo, Instituto de Física, São Paulo, SP, Brazil*

*[18] Universidade Estadual de Campinas, IFGW, Campinas, SP, Brazil*

*[19] Universidade Estadual de Feira de Santana, Brazil*

*[20] Universidade Estadual do Sudoeste da Bahia, Vitoria da Conquista, BA, Brazil*

*[21] Universidade Federal da Bahia, Salvador, BA, Brazil*

*[22] Universidade Federal do ABC, Santo André, SP, Brazil*

*[23] Universidade Federal do Rio de Janeiro, Instituto de Física, Rio de Janeiro, RJ, Brazil*

*[24] Universidade Federal Fluminense, Instituto de Fisica, Niterói, RJ, Brazil*

*[26] Charles University, Faculty of Mathematics and Physics, Institute of Particle and Nuclear Physics, Prague, Czech Republic*

*[27] Institute of Physics of the Academy of Sciences of the Czech Republic, Prague, Czech Republic*

*[28] Palacký University, Olomouc, Czech Republic*

*[30] Institut de Physique Nucléaire d'Orsay (IPNO), Université Paris 11, CNRS-IN2P3, Orsay, France*

*[31] Laboratoire AstroParticule et Cosmologie (APC), Université Paris 7, CNRS-IN2P3, Paris, France*

*[32] Laboratoire de l'Accélérateur Linéaire (LAL), Université Paris 11, CNRS-IN2P3, Orsay, France*

*[33] Laboratoire de Physique Nucléaire et de Hautes Energies (LPNHE), Universités Paris 6 et Paris 7, Paris Cedex 05, France*

[34] Laboratoire de Physique Subatomique et de Cosmologie (LPSC), Université Joseph Fourier, INPG, CNRS-IN2P3, Grenoble, France
[35] SUBATECH, Nantes, France
[36] Bergische Universität Wuppertal, Wuppertal, Germany
[37] Forschungszentrum Karlsruhe, Institut für Kernphysik, Karlsruhe, Germany
[38] Forschungszentrum Karlsruhe, Institut für Prozessdatenverarbeitung und Elektronik, Karlsruhe, Germany
[39] Max-Planck-Institut für Radioastronomie, Bonn, Germany
[40] RWTH Aachen University, III. Physikalisches Institut A, Aachen, Germany
[41] Universität Karlsruhe (TH), Institut für Experimentelle Kernphysik (IEKP), Karlsruhe, Germany
[42] Universität Siegen, Siegen, Germany
[44] Dipartimento di Fisica dell'Università and INFN, Genova, Italy
[45] Università dell'Aquila and INFN, L'Aquila, Italy
[46] Università di Milano and Sezione INFN, Milan, Italy
[47] Dipartimento di Fisica dell'Università del Salento and Sezione INFN, Lecce, Italy
[48] Università di Napoli "Federico II" and Sezione INFN, Napoli, Italy
[49] Università di Roma II "Tor Vergata" and Sezione INFN, Roma, Italy
[50] Università di Catania and Sezione INFN, Catania, Italy
[51] Università di Torino and Sezione INFN, Torino, Italy
[52] Dipartimento di Ingegneria dell'Innovazione dell'Università del Salento and Sezione INFN, Lecce, Italy
[53] Istituto di Astrofisica Spaziale e Fisica Cosmica di Palermo (INAF), Palermo, Italy
[54] Istituto di Fisica dello Spazio Interplanetario (INAF), Università di Torino and Sezione INFN, Torino, Italy
[55] INFN, Laboratori Nazionali del Gran Sasso, Assergi (L'Aquila), Italy
[58] Università di Palermo and Sezione INFN, Catania, Italy
[59] Benemérita Universidad Autónoma de Puebla, Puebla, Mexico
[60] Centro de Investigación y de Estudios Avanzados del IPN (CINVESTAV), México, D.F., Mexico
[61] Instituto Nacional de Astrofisica, Optica y Electronica, Tonantzintla, Puebla, Mexico
[63] Universidad Michoacana de San Nicolas de Hidalgo, Morelia, Michoacan, Mexico
[64] Universidad Nacional Autonoma de Mexico, Mexico, D.F., Mexico
[65] IMAPP, Radboud University, Nijmegen, Netherlands
[66] Kernfysisch Versneller Instituut, University of Groningen, Groningen, Netherlands
[67] NIKHEF, Amsterdam, Netherlands
[68] ASTRON, Dwingeloo, Netherlands
[69] Institute of Nuclear Physics PAN, Krakow, Poland
[70] University of Łódź, Łódź, Poland
[71] LIP and Instituto Superior Técnico, Lisboa, Portugal
[72] J. Stefan Institute, Ljubljana, Slovenia
[73] Laboratory for Astroparticle Physics, University of Nova Gorica, Slovenia
[74] Instituto de Física Corpuscular, CSIC-Universitat de València, Valencia, Spain
[75] Universidad Complutense de Madrid, Madrid, Spain
[76] Universidad de Alcalá, Alcalá de Henares (Madrid), Spain
[77] Universidad de Granada & C.A.F.P.E., Granada, Spain
[78] Universidad de Santiago de Compostela, Spain
[79] Rudolf Peierls Centre for Theoretical Physics, University of Oxford, Oxford, United Kingdom
[81] School of Physics and Astronomy, University of Leeds, United Kingdom
[82] Argonne National Laboratory, Argonne, IL, USA
[83] Case Western Reserve University, Cleveland, OH, USA
[84] Colorado School of Mines, Golden, CO, USA
[85] Colorado State University, Fort Collins, CO, USA
[86] Colorado State University, Pueblo, CO, USA
[87] Fermilab, Batavia, IL, USA
[88] Louisiana State University, Baton Rouge, LA, USA
[89] Michigan Technological University, Houghton, MI, USA
[90] New York University, New York, NY, USA
[91] Northeastern University, Boston, MA, USA
[92] Ohio State University, Columbus, OH, USA
[93] Pennsylvania State University, University Park, PA, USA
[94] Southern University, Baton Rouge, LA, USA
[95] University of California, Los Angeles, CA, USA

*[96] University of Chicago, Enrico Fermi Institute, Chicago, IL, USA*
*[98] University of Hawaii, Honolulu, HI, USA*
*[100] University of Nebraska, Lincoln, NE, USA*
*[101] University of New Mexico, Albuquerque, NM, USA*
*[102] University of Pennsylvania, Philadelphia, PA, USA*
*[103] University of Wisconsin, Madison, WI, USA*
*[104] University of Wisconsin, Milwaukee, WI, USA*
*[105] Institute for Nuclear Science and Technology (INST), Hanoi, Vietnam*
*[‡] Deceased*
*[a] at Konan University, Kobe, Japan*
*[b] On leave of absence at the Instituto Nacional de Astrofisica, Optica y Electronica*
*[c] at Caltech, Pasadena, USA*
*[d] at Hawaii Pacific University*

*Note added: An additional author, C. Hojvat, Fermilab, Batavia, IL, USA, should be added to papers 3,4,5 in this collection*

# Measurement of the cosmic ray energy spectrum above $10^{18}$ eV using the Pierre Auger Observatory

**F. Schüssler* for the Pierre Auger Collaboration†**

* *Karlsruhe Institute of Technology, Karlsruhe, Germany*
† *Observatorio Pierre Auger, Av. San Martin Norte 304, 5613 Malargüe, Argentina*

***Abstract.* The flux of cosmic rays above $10^{18}$ eV has been measured with unprecedented precision using the Pierre Auger Observatory. Two analysis techniques have been used to extend the spectrum downwards from $3 \times 10^{18}$ eV, with the lower energies being explored using a novel technique that exploits the hybrid strengths of the instrument. The systematic uncertainties, and in particular the influence of the energy resolution on the spectral shape, are addressed. The spectrum can be described by a broken power-law of index 3.3 below the ankle which is measured at $\lg(E_{\rm ankle}/{\rm eV}) = 18.6$. Above the ankle the spectrum is described by a power-law $\propto E^{-2.6}$ and a flux suppression with $\lg(E_{1/2}/{\rm eV}) = 19.6$.**

## I. Introduction

Two independent techniques are used at the Pierre Auger Observatory to study extensive air showers created by ultra-high energy cosmic rays in the atmosphere, a ground array of more than 1600 water-Cherenkov detectors and a set of 24 fluorescence telescopes. Construction of the baseline design was completed in June 2008. With stable data taking starting in January 2004, the world's largest dataset of cosmic ray observations has been collected over the last 4 years during the construction phase of the observatory. Here we report on an update with a substantial increase relative to the accumulated exposure of the energy spectrum measurements reported in [1] and [2].

Due to its high duty cycle, the data of the surface detector are sensitive to spectral features at the highest energies. Its energy scale is derived from coincident measurements with the fluorescence detector. A flux suppression around $10^{19.5}$ eV has been established based on these measurements [1] in agreement with the HiRes measurement [3].

An extension to energies below the threshold of $10^{18.5}$ eV is possible with the use of hybrid observations, i.e. measurements with the fluorescence detectors in coincidence with at least one surface detector. Although statistically limited due to the duty-cycle of the fluorescence detectors of about 13%, these measurements make it possible to extend the energy range down to $10^{18}$ eV and can therefore be used to determine the position and shape of the ankle at which the power-law index of the flux changes [4], [5], [6], [7]. A precise measurement of this feature is crucial for an understanding of the underlying phenomena. Several phenomenological models with different predictions and explanations of the shape of the energy spectrum and the cosmic ray mass composition have been proposed [8], [9], [10].

## II. Surface detector data

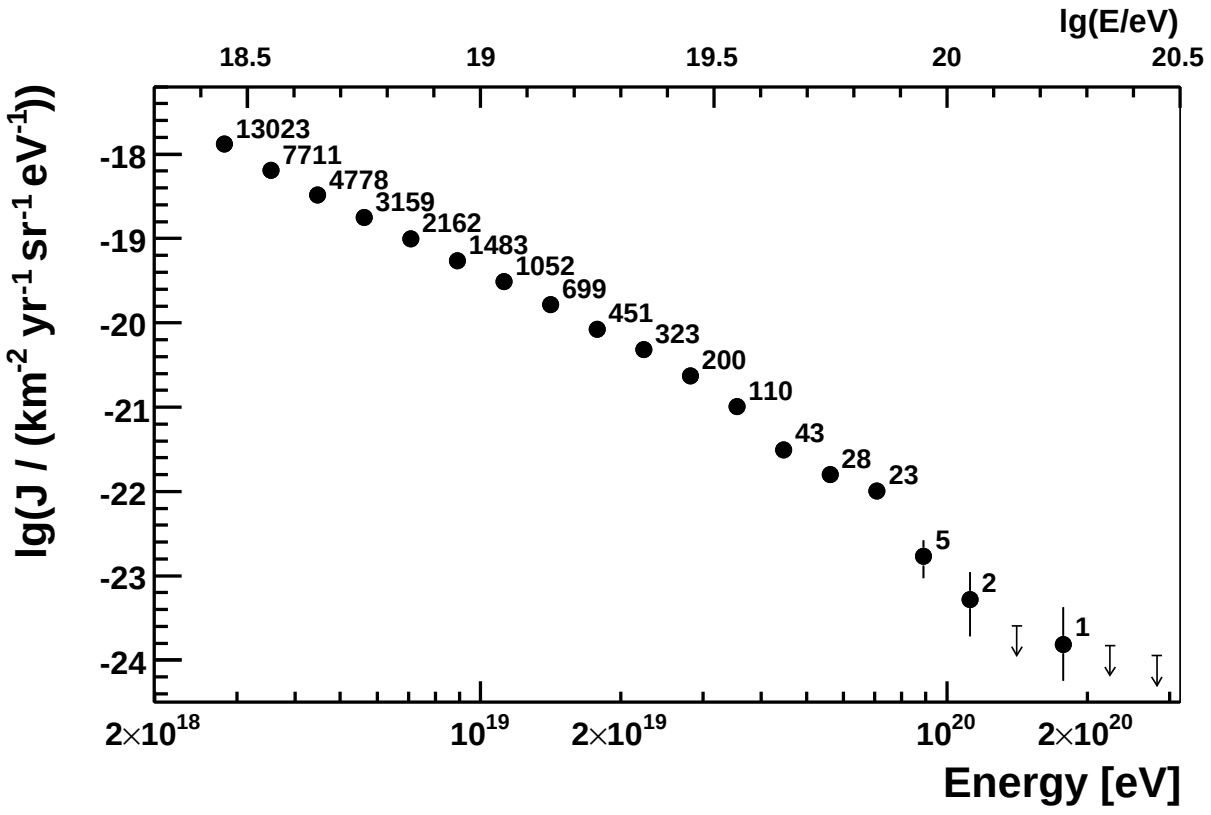

Fig. 1. Energy spectrum derived from surface detector data calibrated with fluorescence measurements. Only statistical uncertainties are shown.

The surface detector array of the Pierre Auger Observatory covers about $3000\,{\rm km}^2$ of the Argentinian Pampa Amarilla. Since its completion in June 2008 the exposure is increased each month by about $350\,{\rm km}^2$ sr yr and amounts to $12,790\,{\rm km}^2$ sr yr for the time period considered for this analysis (01/2004 - 12/2008). The exposure is calculated by integrating the number of active detector stations of the surface array over time. Detailed monitoring information of the status of each surface detector station is stored every second and the exposure is determined with an uncertainty of 3 % [1].

The energy of each shower is calibrated with a subset of high quality events observed by both the surface and the fluorescence detectors after removing attenuation effects by means of a constant-intensity method. The systematic uncertainty of the energy cross-calibration is 7% at $10^{19}$ eV and increases to 15% above $10^{20}$ eV [11].

Due to the energy resolution of the surface detector data of about 20%, bin-to-bin migrations influence the reconstruction of the flux and spectral shape. To correct for these effect, a simple forward- folding approach was applied. It uses MC simulations to determine the energy resolution of the surface detector and derive the bin-to-bin migration matrix. The matrix is then used to derive a flux parameterisation that matches the measured data after forward-folding. The ratio of this parameterisation to the folded flux gives a correction factor that is applied to data. The correction is energy dependent and less than $20\%$ over the full energy range.

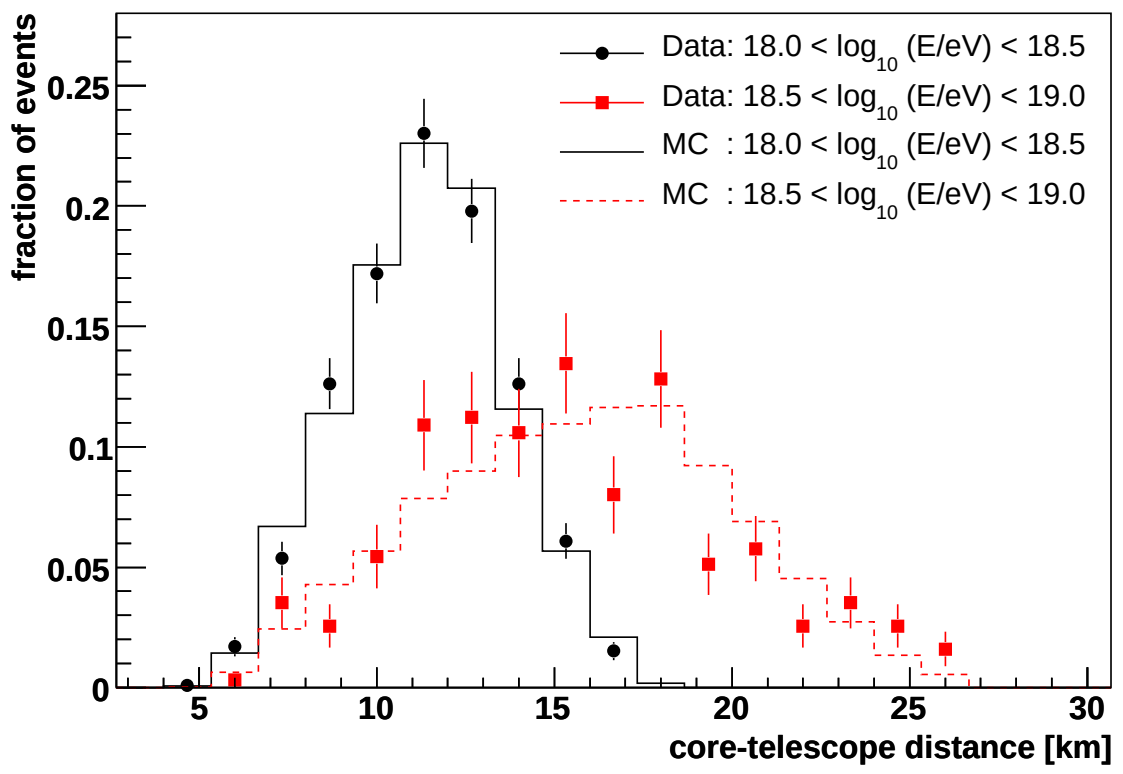

Fig. 2. Comparison between hybrid data and the Monte Carlo simulations used for the determination of the hybrid exposure.

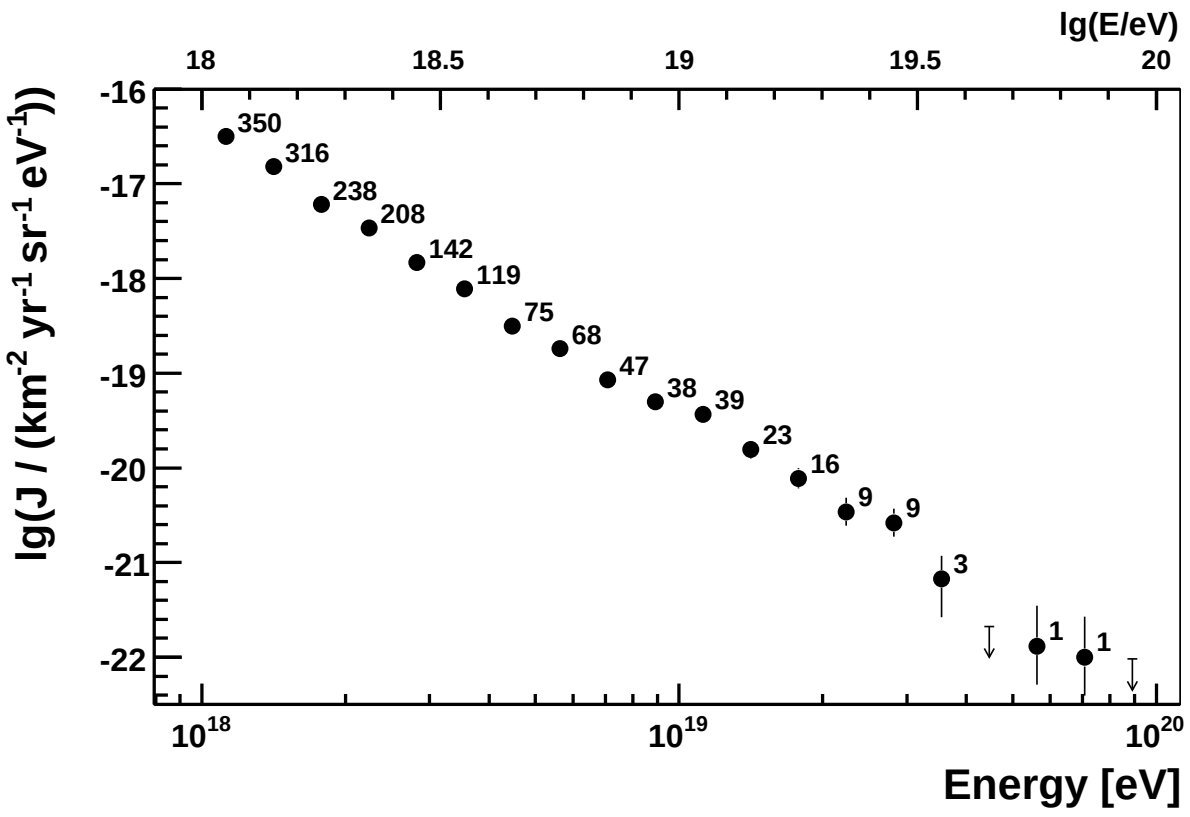

Fig. 3. Energy spectrum derived from hybrid data. Only statistical error bars are shown.

The derived energy spectrum of the surface detector is shown in Fig. 1 together with the event numbers of the underlying raw distribution. Combining the systematic uncertainties of the exposure ($3\%$) and of the forward folding assumptions ($5\%$), the systematic uncertainties of the derived flux is $5.8\%$.

## III. Fluorescence Detector Data

The fluorescence detector of the Pierre Auger Observatory comprises 24 telescopes grouped in 4 buildings on the periphery of the surface array. Air shower observations of the fluorescence detector in coincidence with at least one surface detector permit an independent measurement of the cosmic ray energy spectrum. Due to the lower energy threshold of the fluorescence telescopes, these 'hybrid' events allow us to extend the range of measurement down to $10^{18}$ eV.

The exposure of the hybrid mode of the Pierre Auger Observatory has been derived using a Monte Carlo method which reproduces the actual data conditions of the observatory including their time variability [12]. Based on the extensive monitoring of all detector components [13] a detailed description of the efficiencies of data-taking has been obtained. The time-dependent detector simulation is based on these efficiencies and makes use of the complete description of the atmospheric conditions obtained within the atmospheric monitoring program [14]. For example, we consider only time intervals for which the light attenuation due to aerosols has been measured and for which no clouds have been detected above the observatory [15].

As input to the detector simulation, air showers are simulated with CONEX [16] based on the Sibyll 2.1 [17] and QGSJetII-0.3 [18] hadronic interaction models, assuming a $50\% - 50\%$ mixture of proton and iron primaries. Whereas the derived exposure is independent of the choice of the hadronic interaction model, a systematic uncertainty is induced by the unknown primary mass composition. After applying restrictions to the fiducial volume [19], the systematic uncertainty related to the primary mass composition is $8\%$ at $10^{18}$ eV and becomes negligible above $10^{19}$ eV (see [12] for details).

Additional requirements limit the maximum distance between air shower and the fluorescence detector. They have been derived from comparisons between data and simulated events and assure a saturated trigger efficiency of the fluorescence detector and the independence of the derived flux from the systematic uncertainty of the energy reconstruction. In addition, events are only selected for the determination of the spectrum if they meet certain quality criteria [12], which assure an energy resolution of better than $6\%$ over the full energy range.

Extensive comparisons between simulations and cosmic ray data are performed at all reconstruction levels. An example is the agreement between data and MC in the determination of the fiducial distance shown in Fig. 2. Additional cross-checks involve laser shots fired into the field of view of the fluorescence telescopes from the Central Laser Facility [20]. They have been used to verify the accuracy of the duty cycle.

The design of the Pierre Auger Observatory with its two complementary air shower detection techniques offers the chance to validate the full MC simulation chain and the derived hybrid exposure using air shower observations themselves. Based on this end-to-end verification, the calculated exposure has been corrected by $4\%$. The total systematic uncertainty of the derived hybrid spectrum is $10\%$ at $10^{18}$ eV and decreases to about $6\%$ above $10^{19}$ eV.

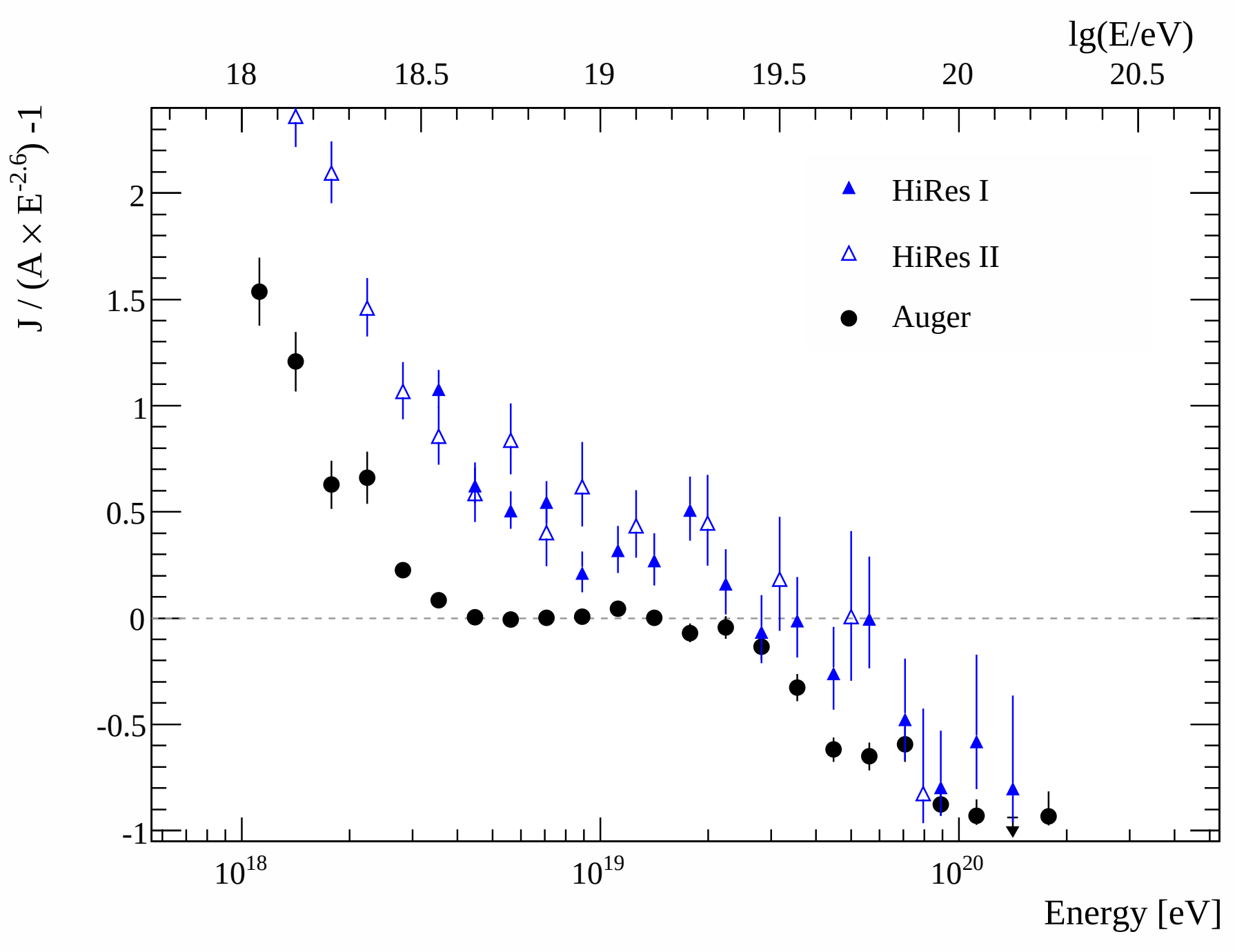

Fig. 4. The fractional difference between the combined energy spectrum of the Pierre Auger Observatory and a spectrum with an index of 2.6. Data from the HiRes instrument [3], [21] are shown for comparison.

The energy spectrum derived from hybrid measurements recorded during the time period 12/2005 - 05/2008 is shown in Fig. 3.

## IV. The combined energy spectrum

The Auger energy spectrum covering the full range from $10^{18}$ eV to above $10^{20}$ eV is derived by combining the two measurements discussed above. The combination procedure utilises a maximum likelihood method which takes into account the systematic and statistical uncertainties of the two spectra. The procedure applied is used to derive flux scale parameters to be applied to the individual spectra. These are $k_{\mathrm{SD}} = 1.01$ and $k_{\mathrm{FD}} = 0.99$ for the surface detector data and hybrid data respectively, showing the good agreement between the independent measurements. The systematic uncertainty of the combined flux is less than $4\%$.

As the surface detector data are calibrated with hybrid events, it should be noted that both spectra share the same systematic uncertainty for the energy assignment. The main contributions to this uncertainty are the absolute fluorescence yield ($14\%$) and the absolute calibration of the fluorescence photodetectors ($9.5\%$). Including a reconstruction uncertainty of about $10\%$ and uncertainties of the atmospheric parameters, an overall systematic uncertainty of the energy scale of $22\%$ has been estimated [11].

The fractional difference of the combined energy spectrum with respect to an assumed flux $\propto E^{-2.6}$ is shown in Fig. 4. Two spectral features are evident: an abrupt change in the spectral index near $4\,\mathrm{EeV}$ (the "ankle") and a more gradual suppression of the flux beyond about $30\,\mathrm{EeV}$.

Some earlier measurements from the HiRes experiment [3], [21] are also shown in Fig. 4 for comparison. A modest systematic energy shift applied to one or both experiments could account for most of the difference between the two. The spectral change at the ankle appears more sharp in our data.

The energy spectrum is fitted with two functions. Both are based on power-laws with the ankle being characterised by a break in the spectral index $\gamma$ at $E_{\mathrm{ankle}}$. The first function is a pure power-law description of the spectrum, i.e. the flux suppression is fitted with a spectral break at $E_{\mathrm{break}}$. The second function uses a smooth transition given by

$$J(E; E > E_{\mathrm{ankle}}) \quad \propto \quad E^{-\gamma_2} \; \frac{1}{1 + \exp\left(\frac{\lg E - \lg E_{1/2}}{\lg W_c}\right)}$$

in addition to the broken power-law to describe the ankle. This fit is shown as black solid line in Fig. 5. The derived parameters (quoting only statistical uncertainties) are:

In Fig. 5 we show a comparison of the combined energy spectrum with spectral shapes expected from different astrophysical scenarios. Assuming for example a uniform distribution of sources, no cosmological evolution of the source luminosity ($(z+1)^m$, i.e. $m = 0$) and a source flux following $\propto E^{-2.6}$ one obtains a spectrum that is at variance with our data. Better agreement is obtained for a scenario including a strong cosmological evolution of the source luminosity ($m = 5$) in combi-nation with a harder injection spectrum ($\propto E^{-2.3}$). A hypothetical model of a pure iron composition injected with a spectrum following $\propto E^{-2.4}$ and uniformly distributed sources with $m = 0$ is able to describe the measured spectrum above the ankle, below which an additional component is required.

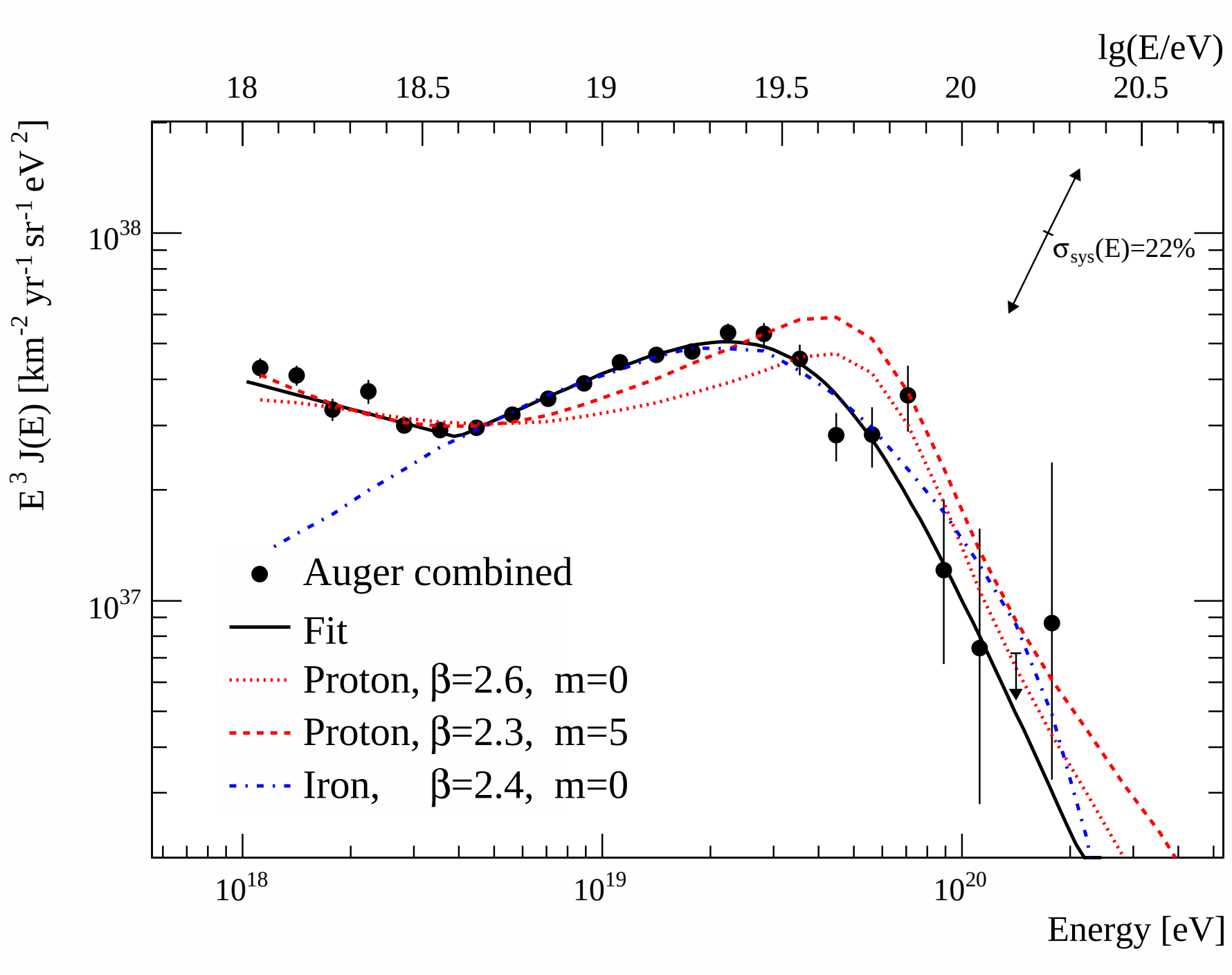

Fig. 5. The combined energy spectrum compared with several astrophysical models assuming a pure composition of protons (red lines) or iron (blue line), a power-law injection spectrum following $E^{-\beta}$ and a maximum energy of $E_{\max} = 10^{20.5}$ eV. The cosmological evolution of the source luminosity is given by $(z+1)^m$. The black line shows the fit used to determine the spectral features (see text). A table with the flux values can be found at [22].

| parameter | broken power laws | power laws + smooth function |
|---|---|---|
| $\gamma_1(E < E_{\text{ankle}})$ | $3.26 \pm 0.04$ | $3.26 \pm 0.04$ |
| $\lg(E_{\text{ankle}}/\text{eV})$ | $18.61 \pm 0.01$ | $18.60 \pm 0.01$ |
| $\gamma_2(E > E_{\text{ankle}})$ | $2.59 \pm 0.02$ | $2.55 \pm 0.04$ |
| $\lg(E_{\text{break}}/\text{eV})$ | $19.46 \pm 0.03$ | |
| $\gamma_3(E > E_{\text{break}})$ | $4.3 \pm 0.2$ | |
| $\lg(E_{1/2}/\text{eV})$ | | $19.61 \pm 0.03$ |
| $\lg(W_c/\text{eV})$ | | $0.16 \pm 0.03$ |

## V. Summary

We presented two independent measurements of the cosmic ray energy spectrum with the Pierre Auger Observatory. Both spectra share the same systematic uncertainties in the energy scale. The combination of the high statistics obtained with the surface detector and the extension to lower energies using hybrid observations enables the precise measurement of both the ankle and the flux suppression at highest energies with unprecedented statistics. First comparisons with astrophysical models have been performed.

# The cosmic ray flux observed at zenith angles larger than 60 degrees with the Pierre Auger Observatory

**R.A. Vázquez * for the Pierre Auger Collaboration**
*University of Santiago de Compostela, Campus Sur s/n, 15782 Santiago de Compostela, Spain
vazquez@fpaxp1.usc.es

***Abstract*. The cosmic ray energy spectrum is obtained using inclined events detected with the surface detectors of the Pierre Auger Observatory. Air showers with zenith angles between 60 and 80 degrees add about 30% to the exposure. Events are identified from background based on compatibility between the arrival time and the detector location enabling the elimination of random signals. The arrival direction is computed using the time information. The core position and a shower size parameter are obtained for each event by fitting measured signals to those obtained from predictions of two-dimensional distributions of the patterns of the muon densities at ground level. The shower size parameter, a zenith angle independent energy estimator, is calibrated using the shower energy measured by the fluorescence technique in a sub-sample of high-quality hybrid events. The measured flux is in agreement with that measured using showers of zenith less than 60 degrees.**

## I. Introduction

Inclined showers are routinely detected by the Pierre Auger Observatory. The Surface Detector (SD) uses 1.2 m deep water–Cherenkov detectors that are sensitive to inclined muons. Hybrid events, events detected simultaneously by the SD and the Fluorescence Detector (FD), provide a method to cross calibrate the Surface Detector even for inclined events.

The analysis of inclined showers is important. It increases the aperture by about 30 % relative to showers with zenith angle less than 60° as used in [12], [10] and has access to regions of the sky which are not visible in the vertical. In addition, inclined showers created by nuclear primaries constitute the background for neutrino detection [5], [15]. Moreover, the inclined showers are characterised by being composed mainly of muons. Therefore they give additional information on the high energy processes in the shower, relevant to the study of composition and of hadronic processes at high energy, see also [16].

Due to the increasing slant depth with the zenith angle, the electromagnetic component is rapidly absorbed as the zenith angle increases. Above $60°$, showers still contain a significant electromagnetic component. For zenith angles larger than 70° the electromagnetic shower is absorbed in the atmosphere and only an electromagnetic 'halo' due to muon decay and other muonic processes survive and account for $\sim 15\%$ of the signal. Due to the long paths traversed, the muons can be deflected by the magnetic field and produce complex patterns at ground where the cylindrical symmetry is lost, depending on the angle between the arrival direction of the shower and the magnetic field. For highly inclined showers ($\geq$ 80°), the magnetic deflection can be so large as to separate the positive and negative muons. This makes the use of the one dimensional lateral distribution functions (LDF), used for zenith angles $< 60°$, unsuitable for analysis of inclined showers. Monte Carlo simulations are used to produce maps of muons arriving at ground. These are either parameterised or kept as histogrammed maps and are used to reconstruct the shower core and a shower size parameter. The electromagnetic component is also parameterised independently using Monte Carlo simulations. Inclined events are reconstructed in a similar manner to the vertical events but taking into account the specific characteristics of inclined showers.

Here we present an update of the analysis of inclined events, in the range from 60° to 80°, in the Pierre Auger Observatory for energies above 6.3 EeV, see also [3], [4].

## II. Event Selection

Events are selected using a chain of quality cuts and triggers, which are similar to the trigger chain used in vertical events [6]. After the single detector triggers, the T3 trigger is the lowest array trigger criterion. Data acquisition distinguishes two types: compact triangles of detectors with long signals and preset patterns of detectors with any signal exceeding a certain threshold. For inclined showers, given their elongated patterns, the more important one is the second, being 63 % of all the events in the 60°–80° zenith angle range. For showers between 70°–80° this fraction increases to 87%.

T3 recorded events are selected at the next trigger level (T4), the physical trigger, if they fulfil a time compatibility test. It is based on a "top down" algorithm where, selected stations are iteratively required to have small time residuals compared to a shower front. In addition a criteria of compactness is also applied. The algorithm is used to iterate over the accepted number of stations until a compatible configuration is obtained.

T4 candidates are reconstructed and their arrival direction, shower size, and the core position are determined. The procedure is described in the next section. For the spectrum analysis high quality events are selected at the next trigger level, the T5 [7], the criteria being that the core must be reconstructed accurately to guarantee a good energy estimation by avoiding events close to the border of the array or events which fall in an area where stations are inactive. Several alternatives were considered. The one currently used (T5HAS) consists of accepting only events were the station closest to the reconstructed core is surrounded by a hexagon of active stations.

The acceptance of the array is then computed geometrically, counting the number of active hexagons, and the aperture is calculated for each array configuration as a function of time. Events with zenith angle greater than $80°$ are not considered in this analysis, as the uncertainty in the angular reconstruction increases with zenith angle, growing rapidly above $80°$. Also at larger zenith angles, due to the low density of muons, the fluctuations are larger and the energy reconstruction has large uncertainty. The total accumulated exposure from 1 January 2004 to 31 December 2008 for zenith angles $< 60°$ is 12790 km$^2$ sr year, the exposure for zenith angles between $60°$ and $80°$ corresponds to 29 % of that value. Over 80000 events were found which pass the T5HAS criteria in the period considered.

## III. Angular and shower size determination

The angular and energy determination of inclined events follows a similar pattern to that for vertical events. For the angular reconstruction the start times of the stations are corrected, taking into account the altitude of the station and the curvature of the Earth (due to the elongated shapes, the shower can spawn several tens of kilometers). The corrected start times are checked against the shower front and the arrival direction is obtained by $\chi^2$ minimisation. We have tested several approaches to the angular reconstruction. In addition, good quality hybrid events can be compared with the Fluorescence Detector reconstruction. Overall, the angular resolution is of the order of $1°$ [8].

For the energy reconstruction the measured signals are compared to the expected ones using the following procedure. First the expected electromagnetic signal, parameterised with Monte Carlo simulations[9], is subtracted from the total signal. At zenith angles $\gtrsim 60°$ the electromagnetic contribution is still appreciable and forms significant fraction of the signal. At larger zenith angles $\gtrsim 70°$, the electromagnetic contribution from $\pi^0$ decay is negligible and only a contribution from the decay of the muons themselves (and other processes) is present. This constitute a fraction of the order of 15 %. After the electromagnetic component has been subtracted, the muonic signal is compared to the expected one taken from 'muon maps'. For inclined events, the lack of cylindrical symmetry around the shower axis makes the use of a single variable LDF impossible. Instead, we have developed muon maps which parameterise the muon number expected as a function of the zenith and azimuth angle. This parameterisation of the muon maps is done in the plane perpendicular to the shower arrival direction. In addition, the response of the detector to inclined muons has been calculated using GEANT4. In the figure 1, we show the probability of muons to produce a given signal for several zenith angles and muon energies. A single muon arriving at $80°$ zenith angle can produce a signal of more than 3 VEM.

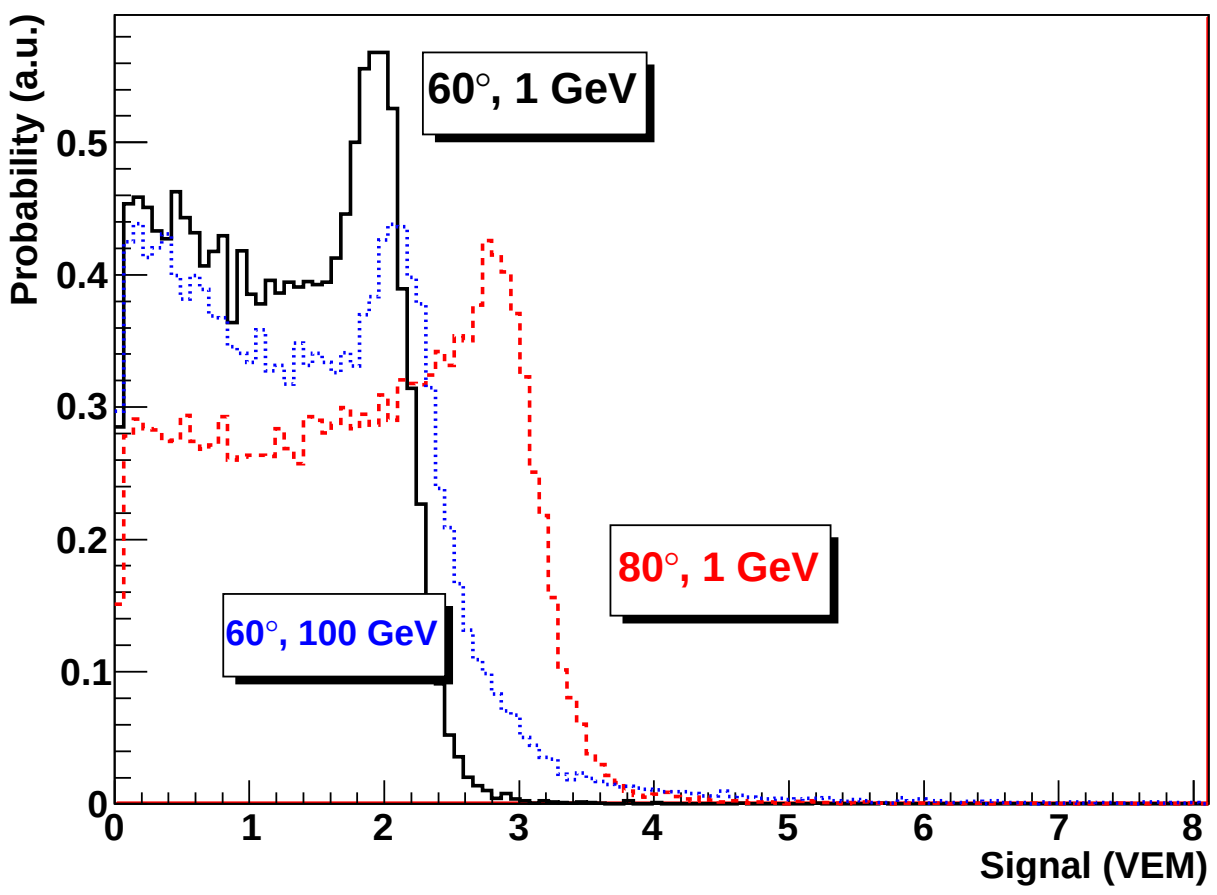

Fig. 1. Surface detector response to inclined muons. Probability of measuring a signal in VEM (vertical equivalent muon) for muons of zenith angle $60°$ and energy 1 GeV (continuous histogram), $60°$ and 100 GeV (dotted histogram), and $80°$ and 1 GeV (dashed histogram).

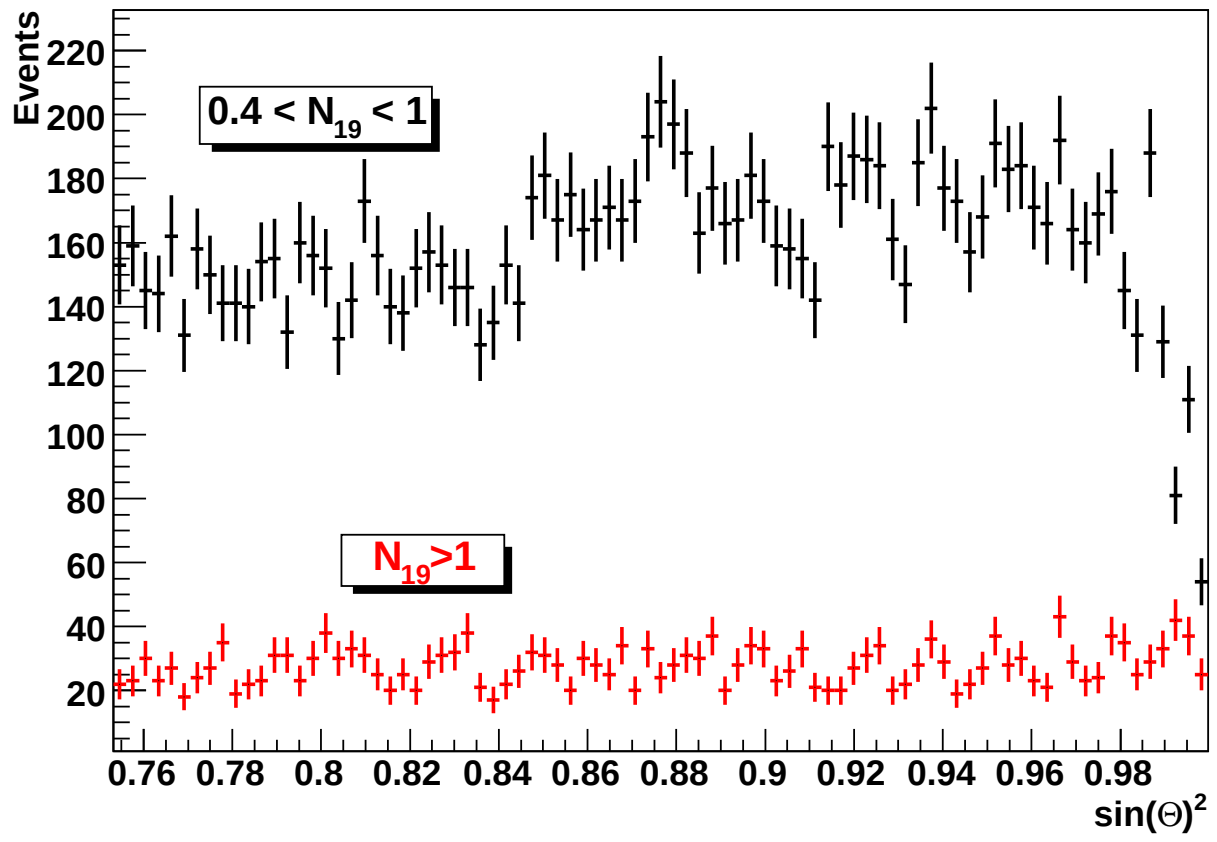

Fig. 2. Distribution of $\sin^2\theta$ for events which pass the T5 trigger and for $N_{19} > 1$ (lower red points) and $0.4 < N_{19} < 1$ (upper black points).

The shower core and the shower size are simultaneously estimated by a likelihood maximisation which accounts for non–triggering and saturated stations. The result of this maximisation procedure is then, the shower size parameter, which can be interpreted as the total number of muons in the shower. From Monte Carlo simulations, it has been found that the number of muons scales with the shower energy and independently of the zenith angle. For convenience, the maps have been normalised by the use of $N_{19}$. $N_{19} = 1$ means that the shower has the same number of muons as a proton shower generated with QGSJET and of energy $10^{19}$ eV.

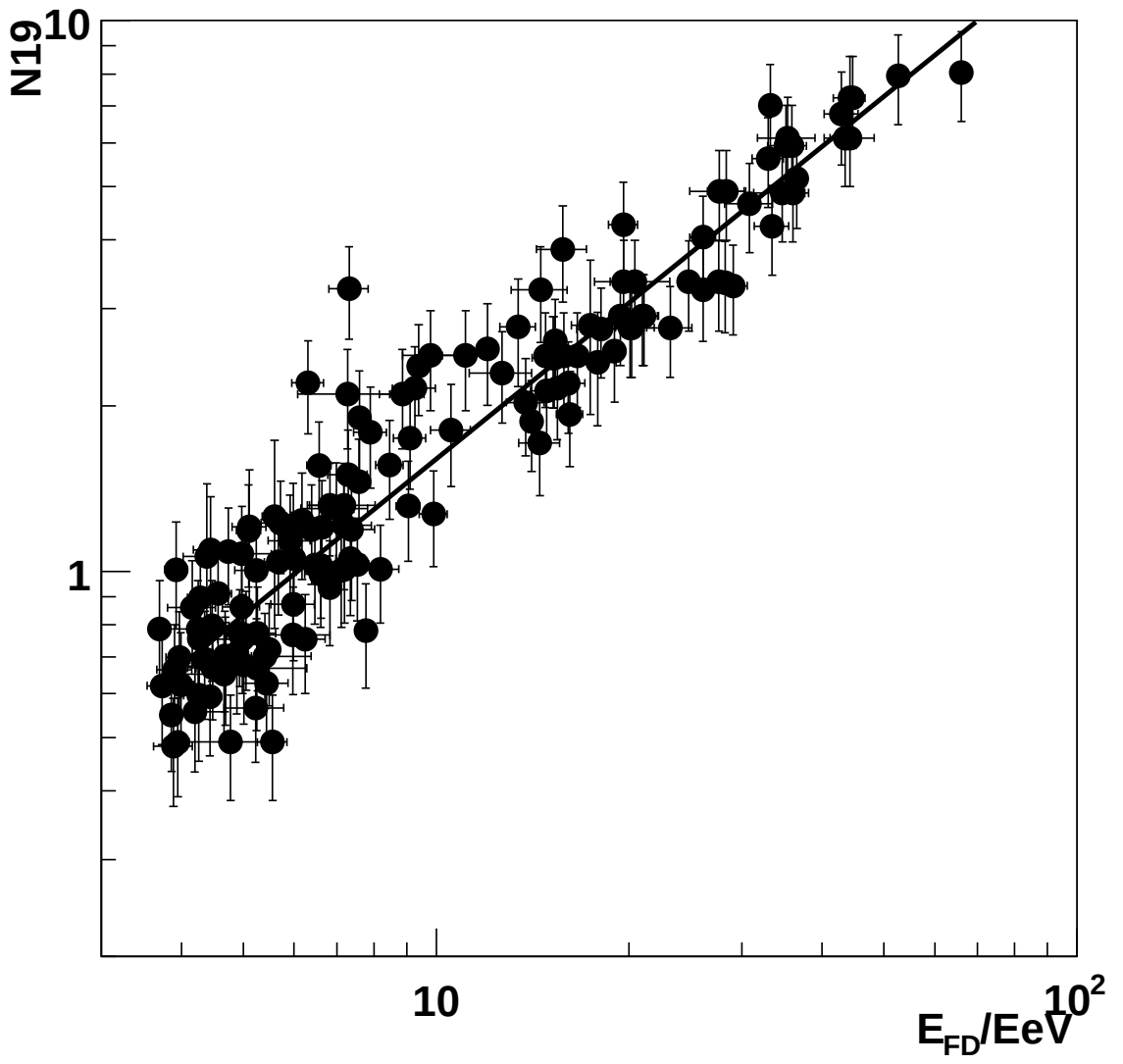

Fig. 3. The $N_{19}$ parameter as a function of the FD energy in EeV. The line is the calibration fit with parameters $a = -0.72 \pm 0.03$ and $b = 0.94 \pm 0.03$, see the text.

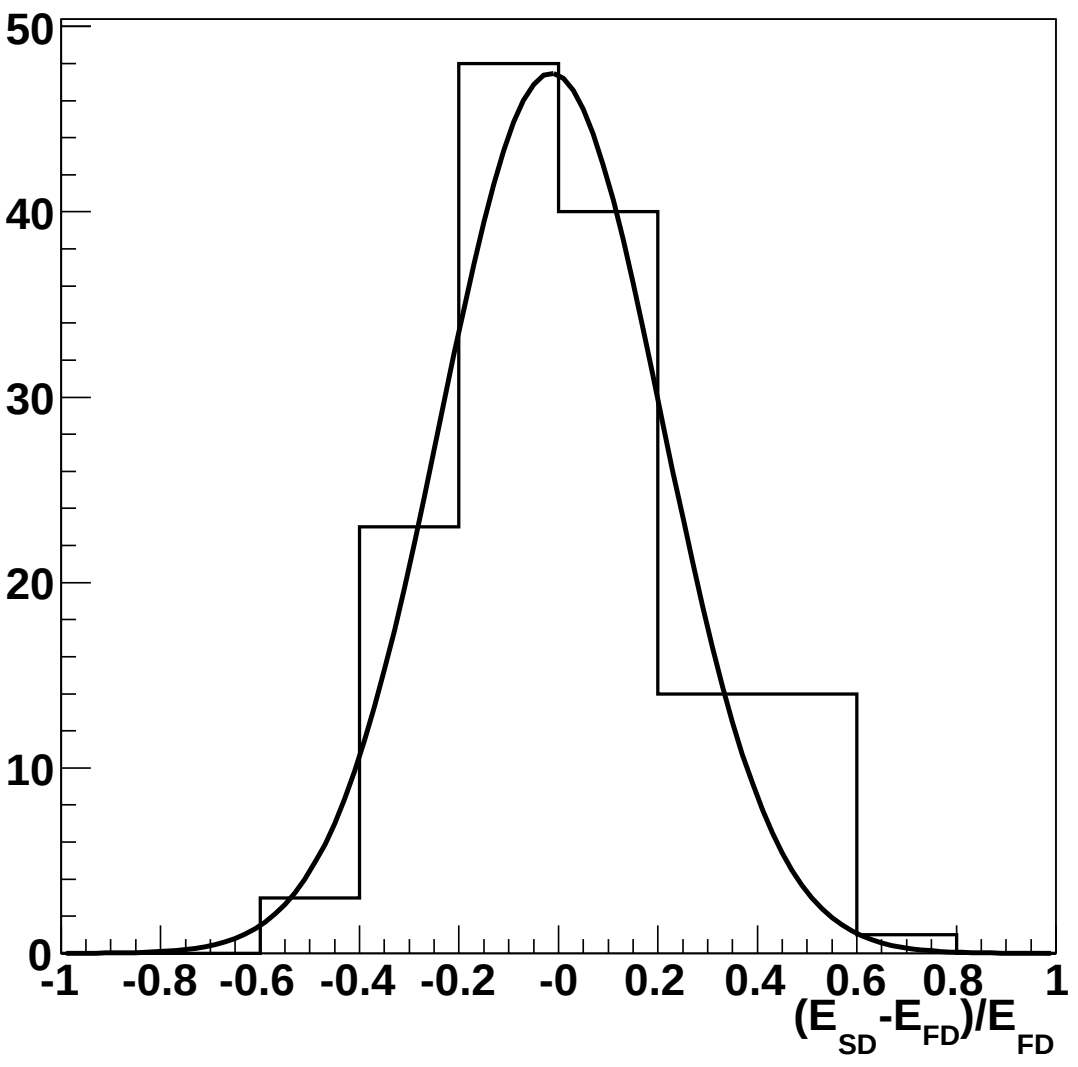

Fig. 4. Relative differences between the FD energy and the calibrated SD energy for events used in the calibration. The line is a Gaussian fit of average 0.01 $\pm 0.02$ and RMS 0.22.

In this way, the zenith angle dependence of the shower size parameter is automatically taken into account. The uncertainty in the determination of $N_{19}$ has been splitted in three terms:

$$\sigma^2_{N_{19}} = \sigma^2_{\rm stat} + \sigma^2_\theta + \sigma^2_{\rm sh}; \qquad (1)$$

where $\sigma_{\rm stat}$ is the statistical uncertainty, obtained from the maximum likelihood, $\sigma_\theta$ is the uncertainty in $N_{19}$ due to the angular reconstruction uncertainty, and $\sigma_{\rm sh}$ is the uncertainty due to the shower–to–shower fluctuations in the number of muons. For the high energy showers considered in this work ($E > 6.3 \times 10^{18}$ eV), $\sigma_{\rm stat} < 10\%$, $\sigma_\theta < 6\%$ and the shower–to–shower fluctuations induce a fluctuation of the order of 18 % in $N_{19}$, making an overall uncertainty of the order of 22 %.

In figure 2, we show the distribution of $\sin^2(\theta)$ of events with $N_{19} > 1$ and $0.4 < N_{19} < 1$. It can be seen that the distribution for $N_{19} > 1$ is flat, showing that the array is fully efficient for $N_{19} > 1$ ($E > 6.3 \times 10^{18}$ eV).

In addition, systematic uncertainties in the determination of $N_{19}$ have been estimated as follows. Several models of the reconstruction procedure are taken into account, including different muon map implementations (generated with Aires and CORSIKA) [1], [13], detector responses, and minimisation procedures. In the present work, two independent reconstruction codes (called A and B) have been used with different muons maps and tank responses. From this, a systematic uncertainty of the order of $20\%$ is obtained for the $N_{19}$. Below, we will show that most of this uncertainty is reabsorbed in the process of calibration.

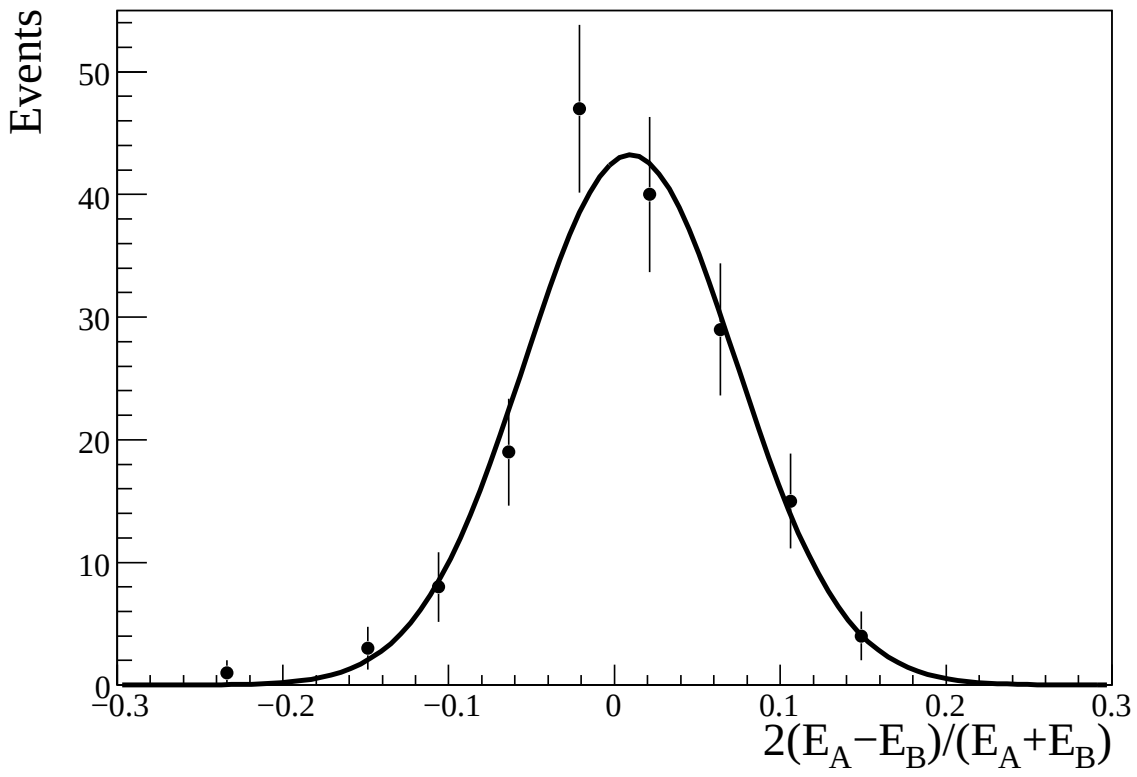

Fig. 5. Relative difference between the energy reconstructed with the two different SD reconstruction procedures $A$ and $B$, as discussed in the text, for events above $E = 10^{19}$ eV. The line is a Gaussian fit to the histogram with reduced $\chi^2$ 5.9/10, mean $0.014 \pm 0.006$ and RMS $0.068 \pm 0.005$.

## IV. Calibration

The absolute energy scale is calibrated using the same procedure adopted in vertical showers, see [14]. A subsample of inclined hybrid events of good quality is selected using a set of cuts [12], optimised for inclined events. For inclined showers, no event above 75° survives the cuts. The energy reconstruction procedure used in the Fluorescence Detector has been described in [11]. Events reconstructed in the SD with $N_{19} < 0.4$ are not considered. This calibration procedure is done independently for the two reconstruction methods discussed earlier. For instance, for the code A, the correlation between the energy obtained from the FD reconstruction and $N_{19}$ is shown in figure 3, where the linear fit $\log_{10}(N_{19}^{A}) = a_A + b_A \log_{10}(E_{\rm FD})$ is also shown. The best fit yields the values $a_A = -0.72 \pm 0.03$ and $b_A = 0.94 \pm 0.03$. In figure 4, we show the relative difference between the energy reconstructed with the Fluorescence Detector and the Surface Detector for these events. A fractional RMS of 22% is found between the two reconstructions, compatible with the estimated uncertainty in the FD reconstruction and the SD reconstruction. The same procedure is applied to the reconstruction code B, obtaining a calibration curve with parameters $a_B = -0.6 \pm 0.01$ and $b_B = 0.93 \pm 0.02$. In the figure 5, we show the relative difference between the two reconstructed energies after the calibration for events above $10^{19}$ eV. The mean difference between the two reconstructed energies is below 2 % and the RMS is of the order of 7%. Therefore, the systematic uncertainty arising from the different reconstruction methods is absorbed in the calibration process, resulting in a systematic uncertainty of the order of 2 %. Other possible sources of systematics are currently under investigation.

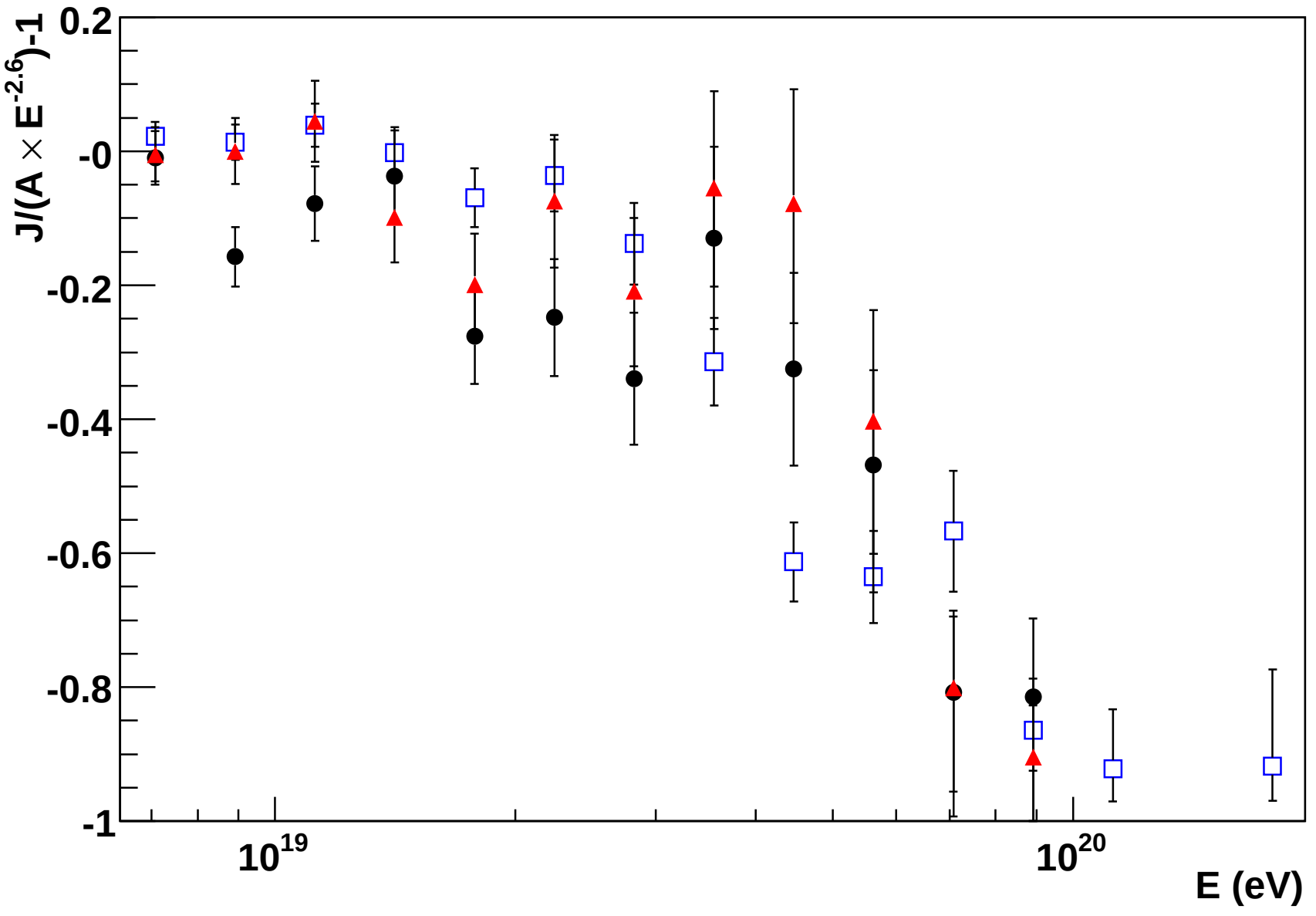

Fig. 6. Fractional differences ($J/(A \times E^{2.6}) - 1$) for the fluxes obtained with the two reconstruction procedures ($A$ (circles) and $B$ (triangles) for inclined events as a function of the energy. Also shown are the raw vertical data [10] (squares).

## V. Results and Discussion

Inclined events recorded from 1 January 2004 to 31 December 2008 were analysed using the procedures outlined above. It was found that above $E = 6.3$ EeV the array is fully efficient to T5HAS triggers (efficiency greater than 98 %). A total of 1750 events where selected above this energy. The fractional difference ($J/(A \times E^{-2.6}) - 1$, where $A$ is a constant) is plotted in figure 6 for the two inclined spectra ($A$ and $B$) and for the raw vertical spectrum supplied by the authors of [10]. At $\log_{10}(E/\mathrm{eV}) < 19.2$ differences between the two inclined spectra are of the order of 10 %. At higher energy, the difference can be as large as 30 %. A power–law fit to the spectra for inclined events gives a slope of $\gamma = 2.79 \pm 0.06$ in the energy range $6.3 \times 10^{18}$ eV to $4.5 \times 10^{19}$ eV. Above $4.5 \times 10^{19}$ eV a power–law fit results in a slope of $\gamma = 5.1 \pm 0.9$. Alternatively, extrapolating the power–law fit with $\gamma = 2.79 \pm 0.06$, we would expect 54 events above $4.5 \times 10^{19}$ eV, where only 39 are observed.

The comparison of the inclined event spectrum to the vertical spectrum can have implications for analysis of composition and of hadronic models. A change on composition or hadronic model would imply a different relation between $N_{19}$ and $E$ to the one used here. This could be seen as a change on the inclined spectrum with respect to the vertical spectrum. This is currently under investigation.

# Energy calibration of data recorded with the surface detectors of the Pierre Auger Observatory

**Claudio Di Giulio***, **for the Pierre Auger Collaboration**†

*Università and INFN di Roma II, ”Tor Vergata”, Via della Ricerca Scientifica 1,00133 Roma, Italy
†Observatorio Pierre Auger, Av. San Martin Norte 304, 5613 Malargüe, Argentinia

***Abstract*. The energy of the primary particles of air showers recorded using the water-Cherenkov detectors of the Pierre Auger Observatory is inferred from simultaneous measurements of showers together with the fluorescence telescopes. The signal on the ground at 1000 $\mathrm{m}$ from the shower axis obtained using the water-Cherenkov detectors is related directly to the calorimetric energy measured with the telescopes. The energy assignment is therefore independent of air-shower simulations except for the assumptions that must be made about the energy carried into the ground by neutrinos and muons. The correlation between the signal at ground and the calorimetric energy is used to derive a calibration curve. A detailed description of the method used to determine the energy scale is presented. The systematic uncertainties on the calibration procedure are discussed.**

## I. Introduction

The Pierre Auger Observatory [1] detects air showers with over 1600 water-Cherenkov detectors, collectively called the surface detector (SD). The SD measures the lateral distribution of particles in air showers with a duty cycle of almost 100% [2]. The SD is overlooked by the fluorescence detector (FD) which consists of 24 fluorescence telescopes grouped in units of 6 at four locations on the periphery of the SD. The FD is only used on clear moonless nights, and has a duty cycle of 13% [3]. The FD provides a nearly calorimetric energy measurement, $E_{\mathrm{FD}}$, since the fluorescence light is produced in proportion to the energy dissipation by a shower in the atmosphere [4], [5]. An example of a reconstruction of a typical air shower with an energy of 40 $\mathrm{EeV}$ and a zenith angle of $36°$ detected with the SD and FD is shown in figures 1 and 2.

The signals recorded in a water-Cherenkov detector are converted in terms of vertical equivalent muons (VEM). One VEM represents the average of the signals produced in the 3 PMTs of the detector by a vertical muon that passes centrally through the SD detector unit. The air shower axis is obtained from the arrival time of the first particles in each detector station. The impact point on ground and the lateral distribution of signals are obtained in a global maximum likelihood minimization which accounts for the station trigger threshold and the overflow of the FADCs counts in the stations very close to the shower axis. The effect of the fluctuation of the lateral distribution function is minimized at $1000\,\mathrm{m}$.

The interpolated signal at a fixed distance from the shower core correlates well with the energy of the primary cosmic ray [6]. The signal at distance of $1000\,\mathrm{m}$, S(1000), indicated as a cross in figure 1 is used as energy estimator.

For the air showers that are also observed with the fluorescence telescopes a direct measurement of the longitudinal profile of the air shower is possible. The longitudinal profile of the air shower, i.e. the energy deposit as a function of traversed matter in the atmosphere, is obtained determining first the shower geometry and then accounting for the fluorescence and Cherenkov light contributions and the light scattering and attenuation [7]. The shower axis is derived using the timing information and the direction of the triggered PMTs of the fluorescence telescope and using the timing information of the water-Cherenkov detector with the highest signal, this allow an angular resolution better than $1°$. The FADCs counts recorded by the PMTs of the fluorescence telescope are converted into photons using the calibration constant derived night by night [8]. The timing information is converted in atmospheric slant depth correcting for the measured atmospheric condition [9]. From the estimated fluorescence light the energy deposit profile is obtained using the absolute fluorescence yield in air which at 293 $\mathrm{K}$ and 1013 $\mathrm{hPa}$ at 337 $\mathrm{nm}$ band is $5.05 \pm 0.71$ photons$/\mathrm{MeV}$ of energy deposited [10]. The fluorescence yield pressure and wavelength dependency are accounted for [11].

Due to the limited field of view, the longitudinal profile is not entirely recorded, so a fit with a Gaisser-Hillas function is employed to obtain the full profile.

The subsample of air showers that are recorded by both detectors, called “hybrid events”, are used to relate $E_{\mathrm{FD}}$ to $S(1000)$. The energy scale inferred from this data sample is applied to the full sample of showers detected by the array of the water-Cherenkov detectors.

## II. Data Analysis

A subset of high-quality hybrid events detected between January 2004 and December 2008 with reconstructed zenith angle less than $60°$ are used in this analysis [12]. To ensure that a shower recorded by the SD has a reliable estimate of $S(1000)$, accidental triggers are rejected and all six nearest neighbours of the station with the largest signal must be active. This guarantees the core of the shower being contained within the array. The geometry of an event is determined from the times recorded at a fluorescence telescope, supplemented by the time of the water-Cherenkov detector with the highest signal. This station must be within $750\,\mathrm{m}$ from the shower axis [13]. The reduced $\chi^2$ of the longitudinal profile fit to the Gaisser-Hillas function [7] has to be less than 2.5. Events are rejected by requiring that the $\chi^2$ of a linear fit to the longitudinal profile exceeds the Gaisser-Hillas fit $\chi^2$ by at least four. The depth of shower maximum, $X_{\max}$, has to be within the field of view of the telescopes and the fraction of the signal detected by the fluorescence telescopes and attributed to Cherenkov light must be less than 50%. The uncertainties on $E_{\mathrm{FD}}$ being lower than 20% and on $X_{\max}$ lower than $40\,\mathrm{g/cm^2}$ are also requested. The selection criteria include a measurement of the vertical aerosol optical depth profile (VAOD(h)) [14] using laser shots generated by the central laser facility (CLF) [15] and observed by the fluorescence telescopes in the same hour of each selected hybrid event. For a given energy the value of $S(1000)$ decreases with zenith angle, $\theta$, due to the attenuation of the shower particles and geometrical effects. Assuming an isotropic flux for the whole energy range considered, we extract the shape of the attenuation curve from the data [16]. The fitted attenuation curve, $CIC(\theta) = 1 + a\ x + b\ x^2$, is a quadratic function of $x = \cos^2\theta - \cos^2 38^\circ$ and is displayed in figure 3 for a particular constant intensity cut which corresponds to $S_{38^\circ} = 47\,\mathrm{VEM}$, with $a = 0.90 \pm 0.05$ and $b = -1.26 \pm 0.21$. The average angle is $\langle\theta\rangle \simeq 38^\circ$ and we take this angle as reference to convert $S(1000)$ into $S_{38^\circ}$ by $S_{38^\circ} \equiv S(1000)/CIC(\theta)$. It may be regarded as the signal $S(1000)$ the shower would have produced had it arrived at $\theta = 38^\circ$.

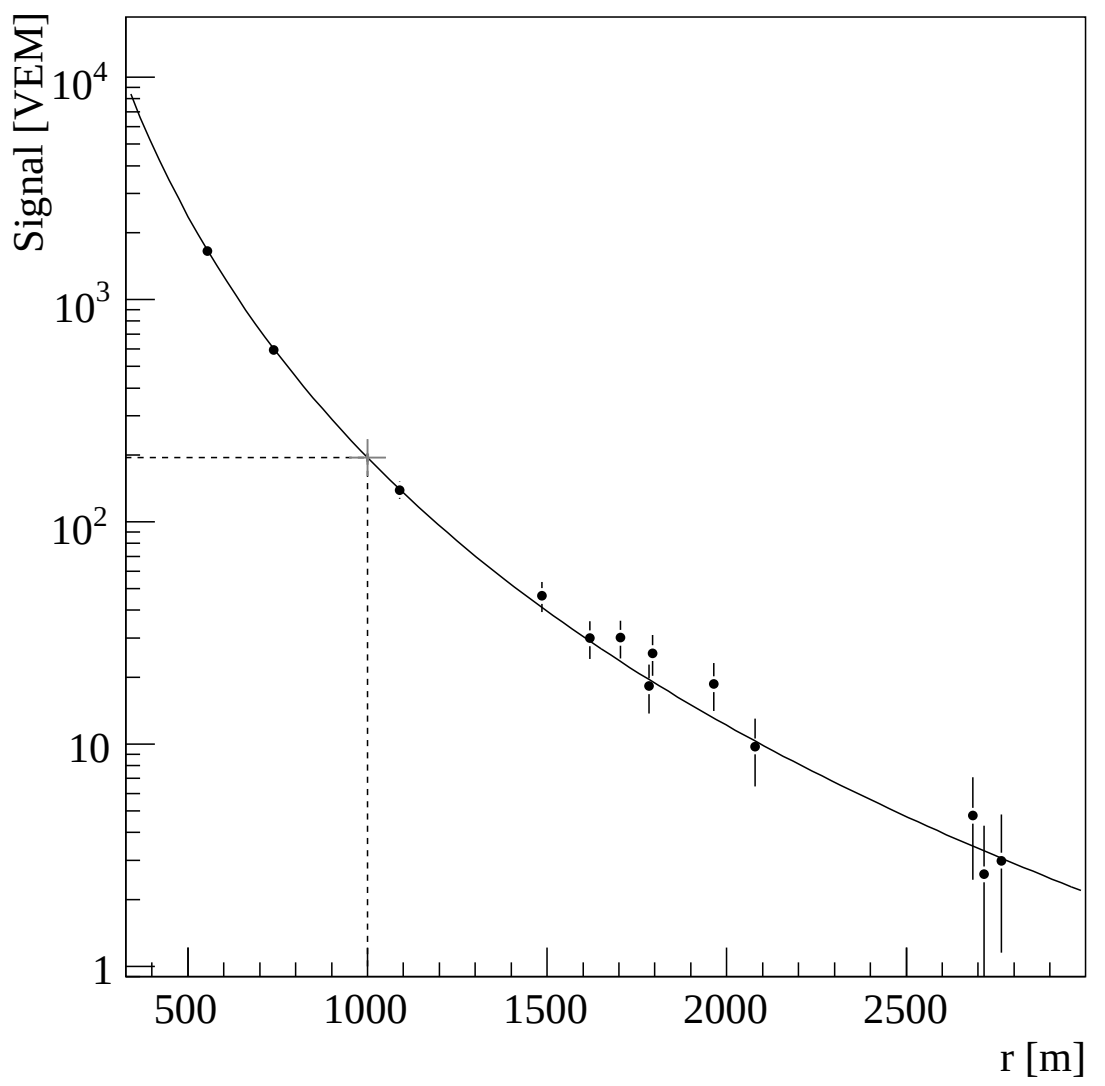

Fig. 1. Lateral distribution: filled circles represent recorded signals. The fitted value S(1000) is marked with a cross.

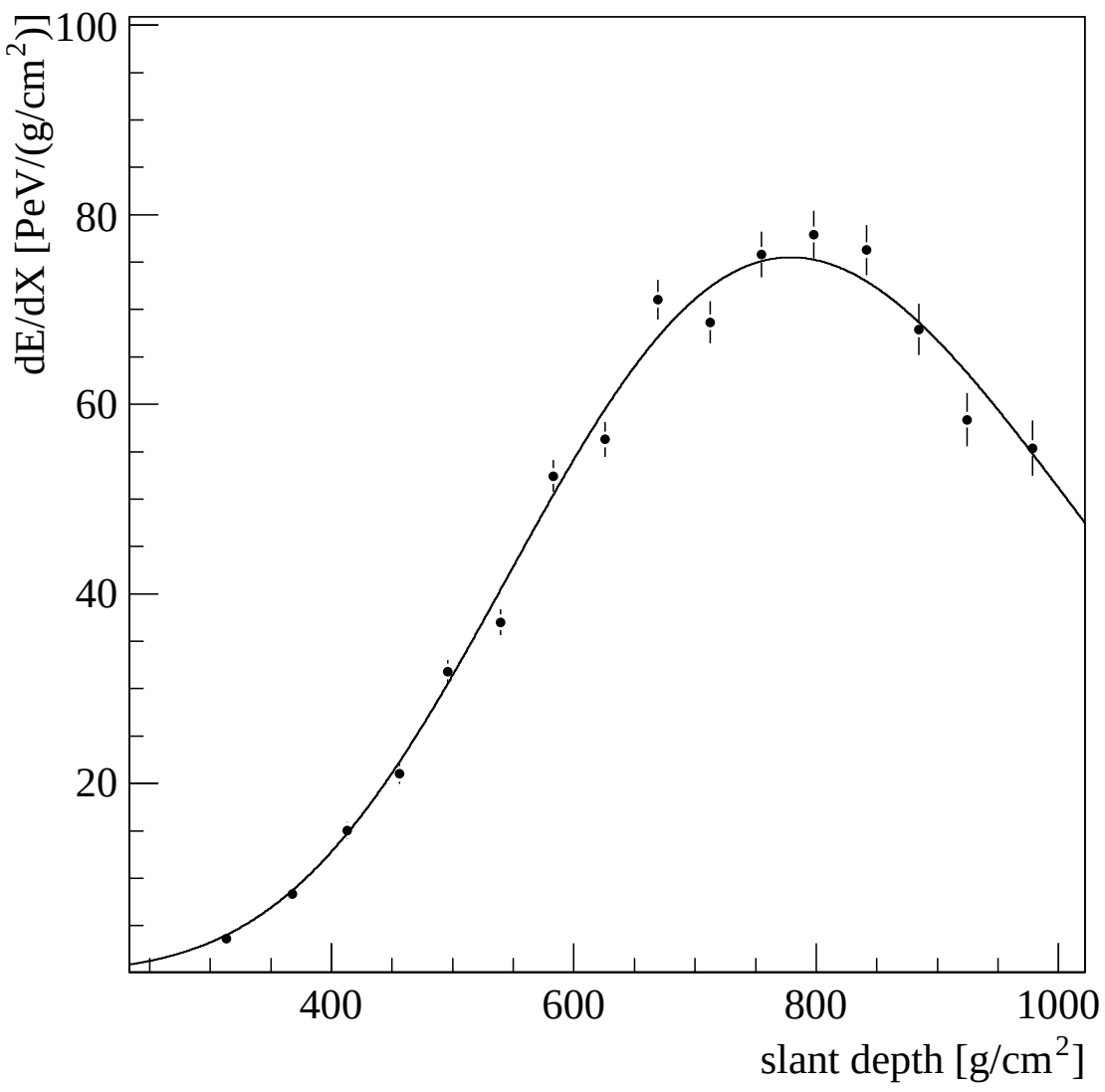

Fig. 2. Longitudinal profile: energy deposit in the atmosphere as a function of the slant depth.

The reconstruction accuracy $\sigma_{S(1000)}$ of the parameter $S(1000)$ is composed by 3 contributions: a statistical uncertainty due to the finite size of the detector and the limited dynamic range of the signal detection, a systematic uncertainty due to the assumptions of the shape of the lateral distribution and finally due to the shower-to-shower fluctuations [17]. The relative uncertainty is shown in figure 4, and in the range of interest, $\sigma_{S_{38^\circ}}/S_{38^\circ} \simeq 14\%$.

Not all the energy of a primary cosmic ray particle ends up in the electromagnetic part of an air shower detected by fluorescence telescopes. Neutrinos escape undetected and muons need long path lengths to release their energy. This non-detected energy is sometimes called the *invisible energy*, and is usually accounted for by correcting the electromagnetic energy $E_{\rm em}$, detected by fluorescence telescopes. The factor $f_{inv}$ is determined from shower simulations to obtain the total shower energy $E_{\rm FD} = f_{inv}\ E_{\rm em}$. The *invisible energy* correction is based on the average for proton and iron showers simulated with the QGSJet model and sums up to about 10% at $10\ EeV$. The neutrino and muon production probabilities have energy dependencies due to the meson decay probabilities in the atmosphere. Thus, the factor $f_{inv}$ depends on the energy for different hadronic interaction models and is also subject to shower-to-shower fluctuations [18].

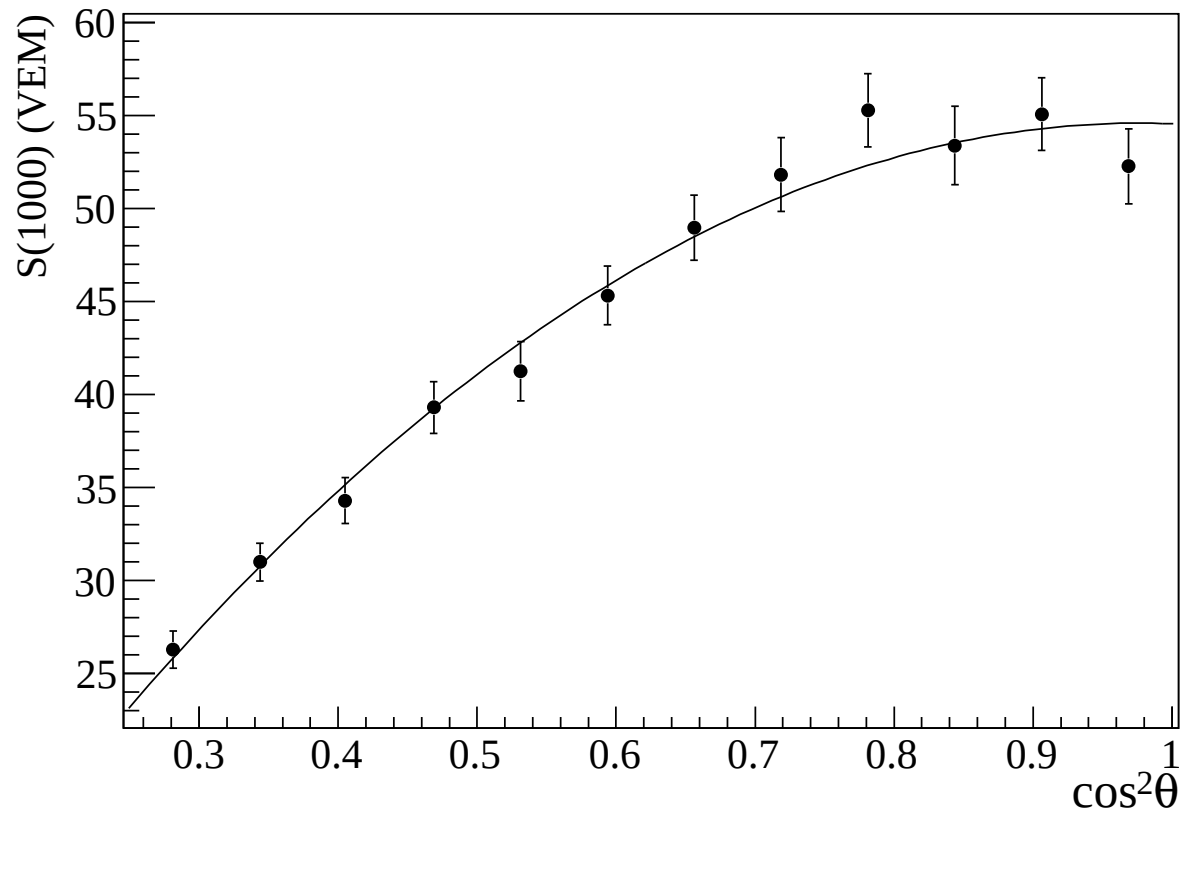

Fig. 3. Derived attenuation curve, $CIC(\theta)$, fitted with a quadratic function.

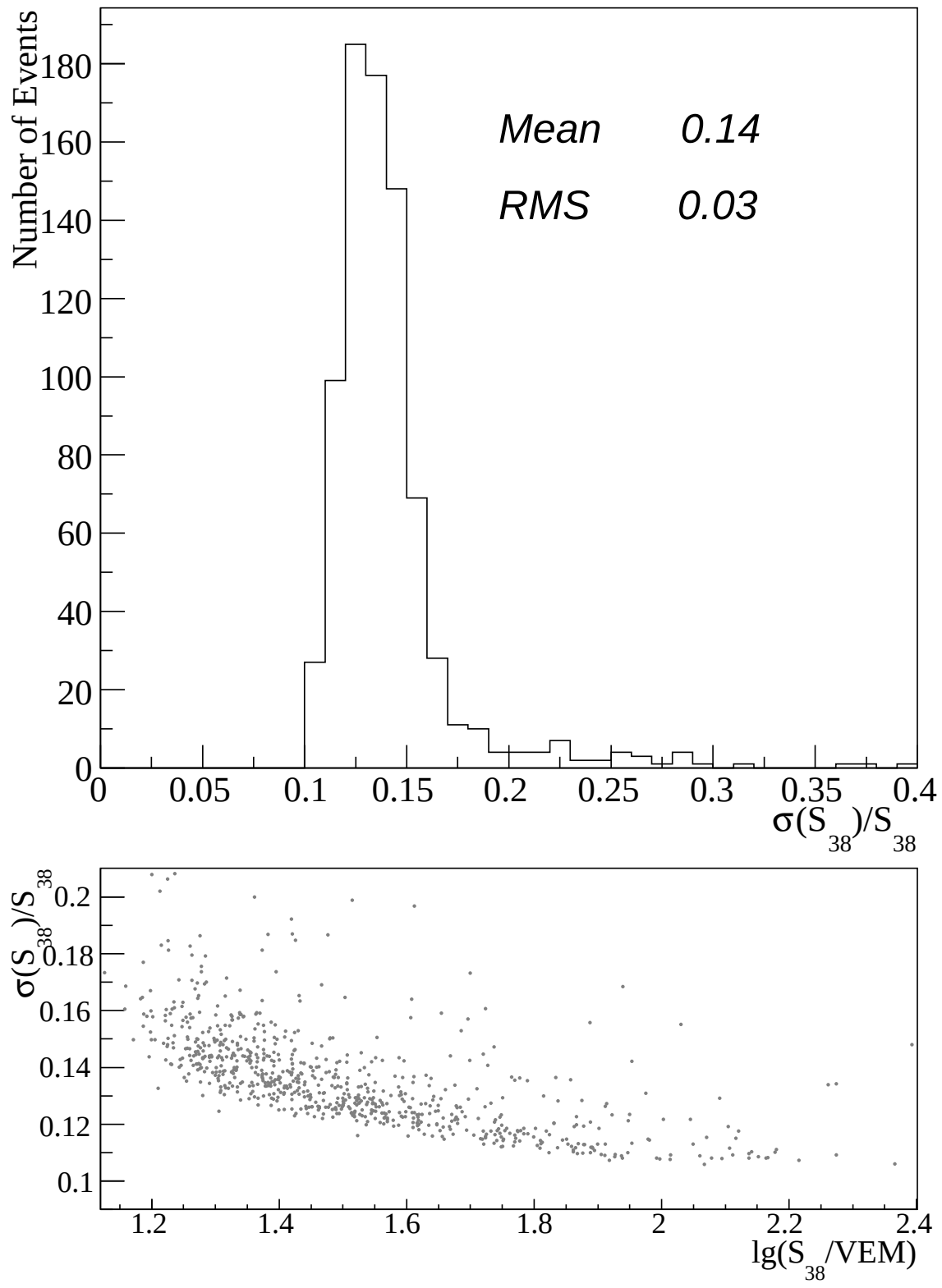

Fig. 4. Upper panel: $S_{38^\circ}$ resolution. Lower panel $\sigma_{S_{38^\circ}}/S_{38^\circ}$ on function of $lg(S_{38^\circ}/VEM)$ scatter plot.

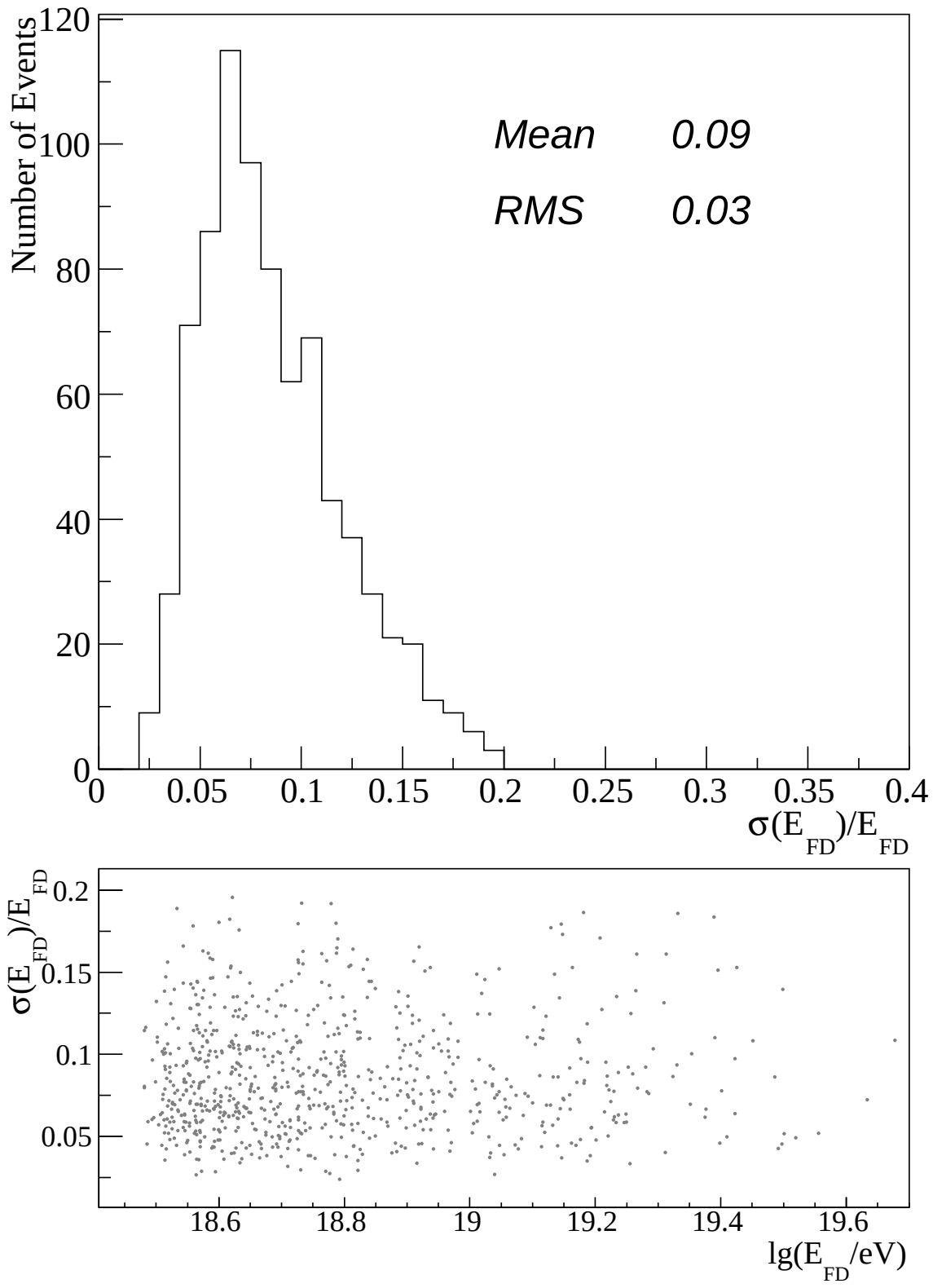

Fig. 5. Upper panel: $E_{\rm FD}$ resolution. Lower panel $\sigma_{E_{\rm FD}}/E_{\rm FD}$ on function of $lg(E_{\rm FD}/eV)$ scatter plot.

The statistical uncertainties, $\sigma_{E_{\rm FD}}$, of the total energy, $E_{\rm FD}$, measured by the fluorescence telescopes is composed of the statistical uncertainty of the light flux, $\sigma_{flux}$, the uncertainty due to the core location and shower direction, $\sigma_{geo}$, the uncertainty on the invisible energy correction, $\sigma_{inv}$ and the uncertainty related to the measured VAOD profile, $\sigma_{atm}$. The total relative uncertainty is about $\sigma_{E_{\rm FD}}/E_{\rm FD} = 9\%$ as shown in figure 5 and does not depend strongly on the energy.

## III. Calibration Curve

The relation of $S_{38}$ and $E_{\rm FD}$ for the 795 hybrid selected events in the energy region where the surface detector array is fully efficient, $E \geq 3\ EeV$, is well described by a power-law function,

$$E = a\ S_{38}^{b}, \tag{1}$$

as shown in figure 6. The results of a fit to the data are

$$a = (1.51 \pm 0.06(stat) \pm 0.12(syst)) \times\ 10^{17}\,{\rm eV},$$
$$b = 1.07 \pm 0.01(stat) \pm 0.04(syst),$$

with a reduced $\chi^2$ of 1.01. $S_{38}$ grows approximately linearly with energy. The root-mean-square deviation of the distribution is about 17% as shown in figure 7, in good agreement with the quadratic sum of the statistical uncertainties of $S_{38^\circ}$ and $E_{\rm FD}$. The calibration accuracy at the highest energies is limited by the number of recorded showers: the most energetic selected event is about $6 \times 10^{19}$ eV. The calibration at low energies extends below the range of interest.

## IV. Systematic Uncertainties

The systematic uncertainty due to the calibration procedure is 7% at $10^{19}$ eV and 15% at $10^{20}$ eV.

The systematic uncertainties on the energy scale $E_{\rm FD}$ sum up to 22%. The largest uncertainties are given by the absolute fluorescence yield (14%) [10], the absolute calibration of the fluorescence telescopes (9%) and the uncertainty due to the reconstruction method of the longitudinal shower profile (10%).

The uncertainty due to the water vapour quenching on the fluorescence yield (5%) is taken into account as described in [19]. Additionally, the wavelength dependent response of the fluorescence telescopes (3%), the uncertainties on measurements of the molecular optical depth (1%), on the measurements of the aerosol optical depth (7%) and on multiple scattering models (1%) are included in the overall systematic uncertainty. The *invisible energy* correction contributes 4% to the total systematic uncertainty of 22% [20].

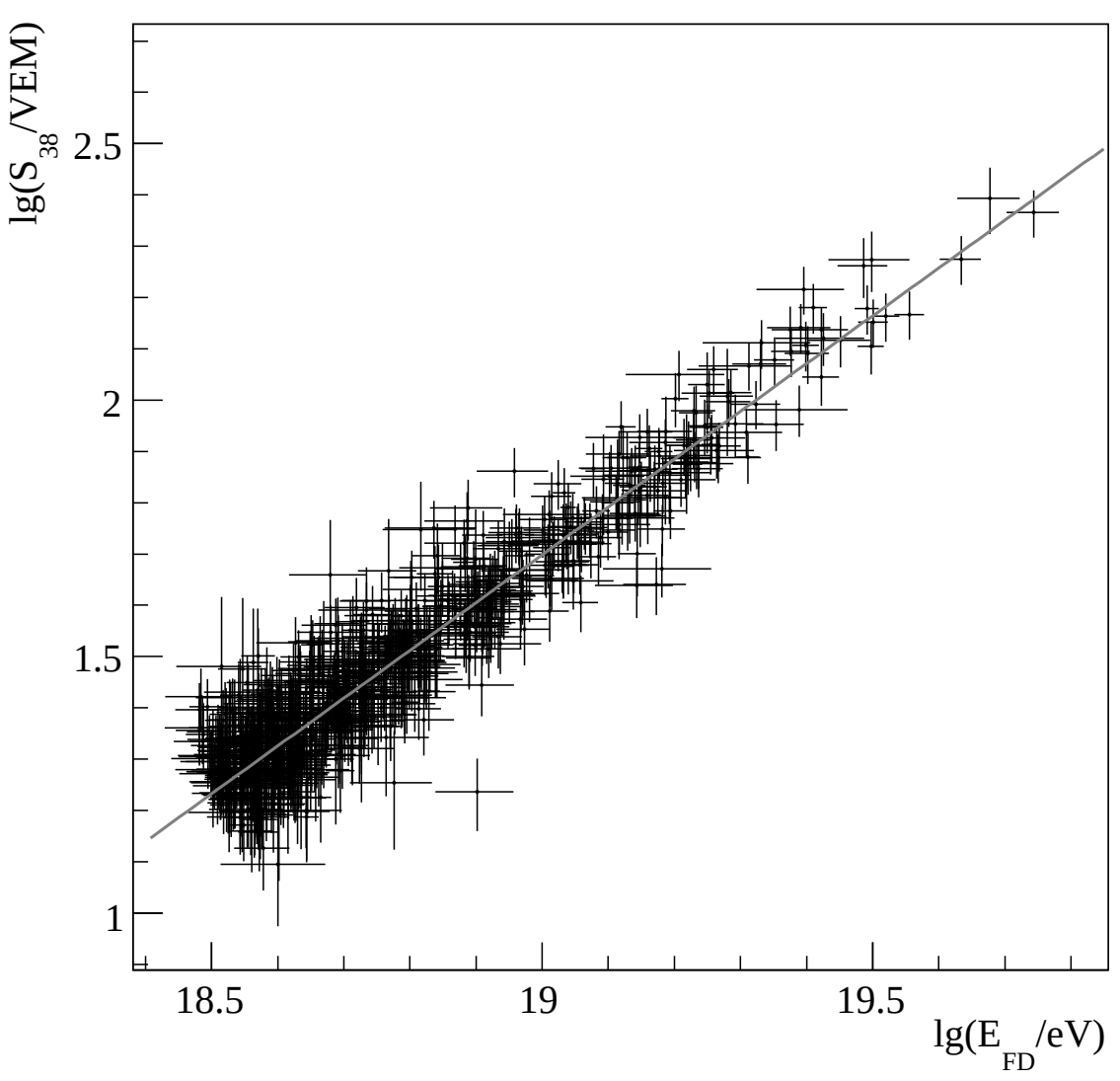

Fig. 6. Correlation between $\lg S_{38}$ and $\lg E_{\rm FD}$ for the 795 hybrid events used in the fit. The line represents the best fit.

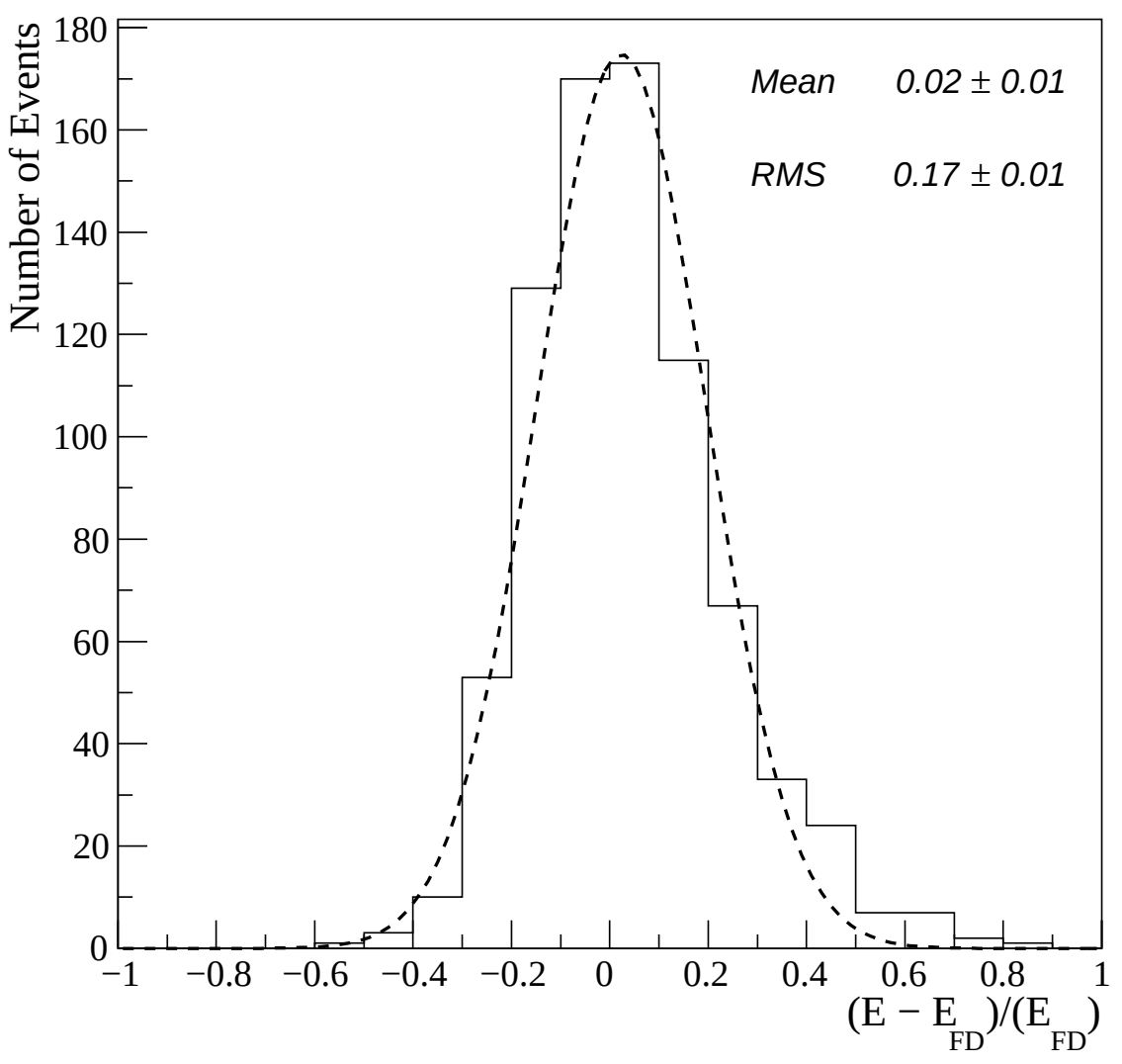

Fig. 7. Fractional difference between the calorimetric energy, $E_{\rm FD}$, and the energy estimate of the surface detector, E, obtained by the calibration curve, for the 795 selected events.

## V. Outlook

The energy calibration of the surface detector array was obtained with measurements of the fluorescence telescopes and a detailed study of the uncertainties was given. Several activities are on-going to reduce the systematic uncertainties of the energy estimate, e.g. the longitudinal profile reconstruction method and the uncertainty of the fluorescence yield. The spectrum derived from data of the surface detector array is calibrated using the method presented in this paper and compared with a spectrum based on measured hybrid data in [21].

# Exposure of the Hybrid Detector of The Pierre Auger Observatory

**Francesco Salamida**$^{*\dagger}$ **for the Pierre Auger Collaboration**

$^*$*University of L'Aquila & INFN, via Vetoio I, 67100, Coppito(AQ)*
$^\dagger$*Observatorio Pierre Auger, Av. San Martín Norte 304, Malargüe, (5613) Mendoza, Argentina*

***Abstract*. The exposure of the Pierre Auger Observatory for events observed by the fluorescence detector in coincidence with at least one station of the surface detector is calculated. All relevant monitoring data collected during the operation, like the status of the detector, background light and atmospheric conditions are considered in both simulation and reconstruction. This allows to realistically reproduce time dependent data taking conditions and efficiencies.**

## I. Introduction

The measurement of the cosmic ray flux above $10^{18}$eV is one of the foremost goals of the Pierre Auger Observatory [1]. In this energy region two different features, the *ankle* and the *GZK cut-off* are expected. In particular the transition between the galactic and the extragalactic component of cosmic rays [2] is widely believed to be associated with a flattening of cosmic rays energy spectrum, identified as the ankle.An accurate determination of the ankle could help to discriminate among theoretical models [3], [4], [5] describing this transition.

The hybrid approach is based on the detection of showers observed by the Fluorescence Detector (FD) in coincidence with at least one station of the Surface Detector (SD). Although a signal in a single station doesn't ensure an independent trigger and reconstruction in SD [6], it is a sufficient condition for a very accurate determination of the shower geometry using the hybrid reconstruction.

The measurement of cosmic ray flux relies on the precise determination of detector exposure that is influenced by several factors. The response of the hybrid detector is in fact very much dependent on energy, distance of recorded event, atmospheric and data taking conditions.

## II. Hybrid Exposure

The flux of cosmic rays $J$ as a function of energy is defined as:

$$J(E) = \frac{\mathrm{d}^4 N}{\mathrm{d}E\ \mathrm{d}S\ \mathrm{d}\Omega\ \mathrm{d}t} \simeq \frac{1}{\Delta E}\frac{N^D(E)}{\mathcal{E}(E)}; \qquad (1)$$

where $N^D(E)$ is the number of detected events in the energy bin centered around $E$ having width $\Delta E$ on a surface element d$S$, solid angle d$\Omega$ and time d$t$, $\mathcal{E}(E)$ represents the energy dependent exposure of the detector.

The exposure, as a function of primary shower energy, can be written as:

$$\mathcal{E}(E) = \int_T \int_\Omega \int_{A_{gen}} \varepsilon(E, t, \theta, \phi)\ \mathrm{d}S \cos\theta\ \mathrm{d}\Omega\ \mathrm{d}t; \qquad (2)$$

where $\varepsilon$ is the detection efficiency including analysis selection cuts, d$S$ and A$_{gen}$ are the differential and total Monte Carlo generation areas, respectively. The generation area A$_{gen}$ has been chosen large enough to include the detector array with a sizable trigger efficiency. $\mathrm{d}\Omega = \sin\theta \mathrm{d}\theta \mathrm{d}\phi$ and $\Omega$ are respectively the differential and total solid angles, being $\theta$ and $\phi$ the zenith and azimuth angles. The growth of surface array and ongoing extensions of the fluorescence detector, seasonal and instrumental effects obviously introduce changes of the detector configuration with time. All these effects can be taken into account by simulating a sample of events that reproduces the exact data taking conditions.

## III. Hybrid On-Time

The calculation of hybrid exposure requires the knowledge of the detector on-time. The efficiency of fluorescence and hybrid data taking is influenced by many effects. These can be external, e.g. lightnings and storms, or internal, due to data taking inefficiencies, e.g. DAQ failures. To determine the on-time of our hybrid detector it is therefore crucial to take as many of these possibilities into account and derive a solid description of the time dependent data taking.

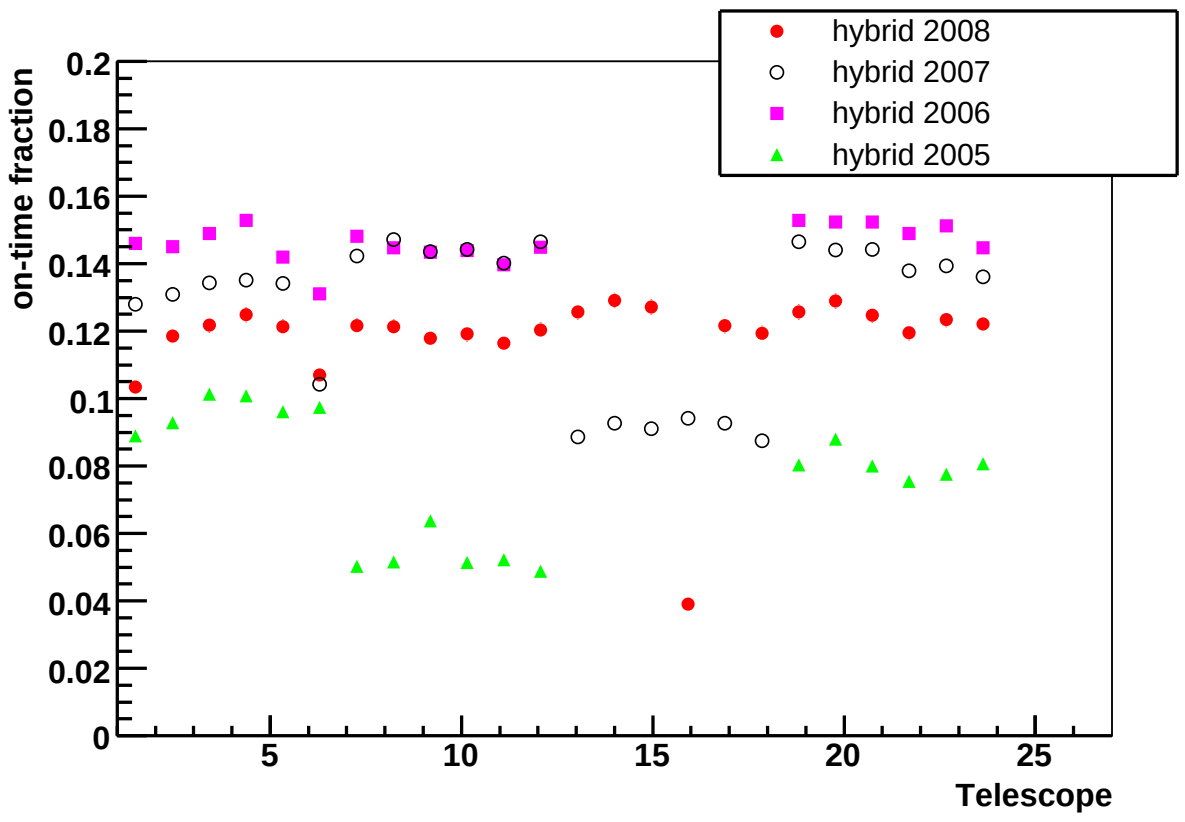

Fig. 1. The evolution of the average hybrid duty-cycle during the construction phase of the Pierre Auger Observatory.

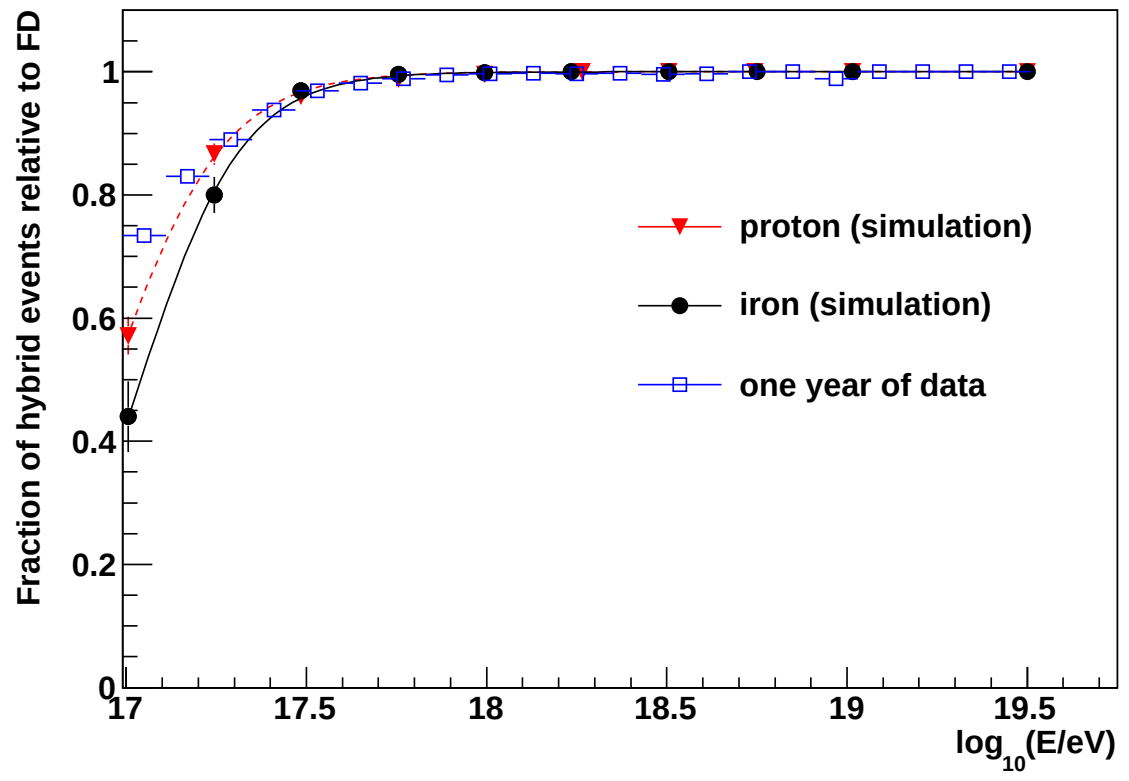

Fig. 2. Relative hybrid trigger efficiency from hybrid simulation for proton, iron and data. All the events are taken for zenith less than 60°.

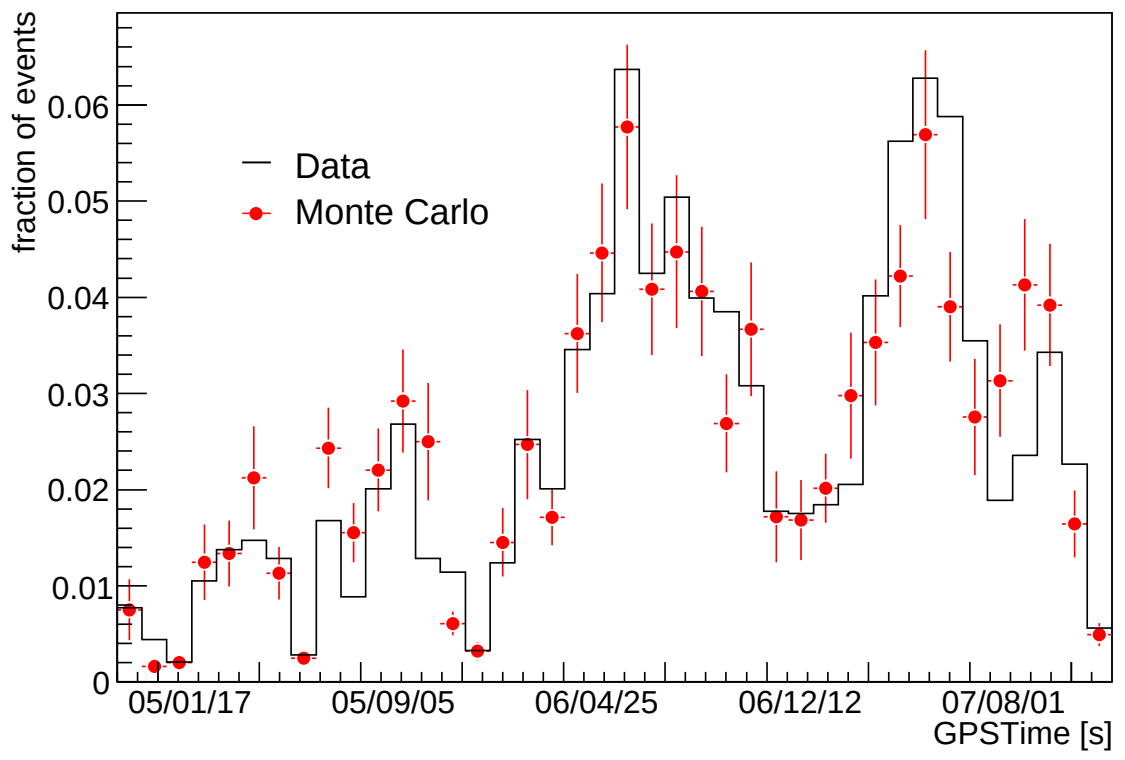

Fig. 3. Data-Monte Carlo Comparison: fraction of hybrid events as a function of time starting from November 2005. Both data (solid line) and simulations (solid circles) are shown.

Failures can occur on different levels starting from the smallest unit of the FD, i.e. one single PMT readout channel, up to the highest level, i.e. the combined SD-FD data taking of the Observatory.

The on-time of the hybrid detector has been derived using data and a variety of monitoring information. As compromise between accuracy and stability we derived the complete detector status down to the single photo-multiplier for time intervals of $10$ min.

The time evolution of the full hybrid duty-cycle over 4 years during the construction phase of the observatory is given in figure 1. It has to be noted that the telescopes belonging to the building of Los Morados (telescopes 7-12) have become operational only in May 2005 and the ones in Loma Amarilla (telescopes 13-18) in March 2007. Moreover the quality of the data taking increases from 2005 to 2007. The decrease of the average on-time in 2008 is due to the lowering of the maximum background value allowed for the FD data taking. The result has been cross-checked with other independent analyses [7], [8] giving an overall agreement within about 4%.

## IV. Monte Carlo simulation and Event Selection

To reproduce the exact working conditions of the experiment and the entire sequence of the different occurring configurations, a large sample of Monte Carlo events has been produced. The simulated data sample consists of longitudinal energy deposit profiles generated with the CONEX [12] code using QGSJet-II [10] and Sibyll 2.1 [11] as hadronic interaction models. As the distribution of particles at ground is not provided by CONEX, the time of the station with the highest signal is simulated according to the muon arrival time distribution [13]. This time is needed in the hybrid reconstruction for determining the incoming direction of the showers and the impact point at ground.

In order to validate such approach, a full hybrid simulation was performed using CORSIKA showers [15], in which FD and SD response are simultaneously and fully simulated. As it is shown in Figure 2, the hybrid trigger efficiency (an FD event in coincidence with at least one SD station) is flat and equal to 1 at energies greater than $10^{18}$ eV. The difference between the two primaries becomes negligible for energies larger than $10^{17.5}$ eV. Furthermore the comparison with data shows a satisfactory agreement. The CORSIKA simulations have been also used to parameterize the response of the SD stations using the *Lateral Trigger Probability* functions [16].

The effect of the different data taking configurations has been taken into account and simulated using the calculation of the hybrid detector on-time. Moreover the impact of cloud coverage and atmospheric conditions on the exposure calculation has been taken into account using the information of the atmospheric monitoring [9] of the Pierre Auger Observatory. All the simulations were performed within the Auger analysis framework [14].

Once the shower geometry is known, the longitudinal profile can be reconstructed and the energy calculated in the same way as for data. The following quality cuts have been designed and used also for the hybrid spectrum.

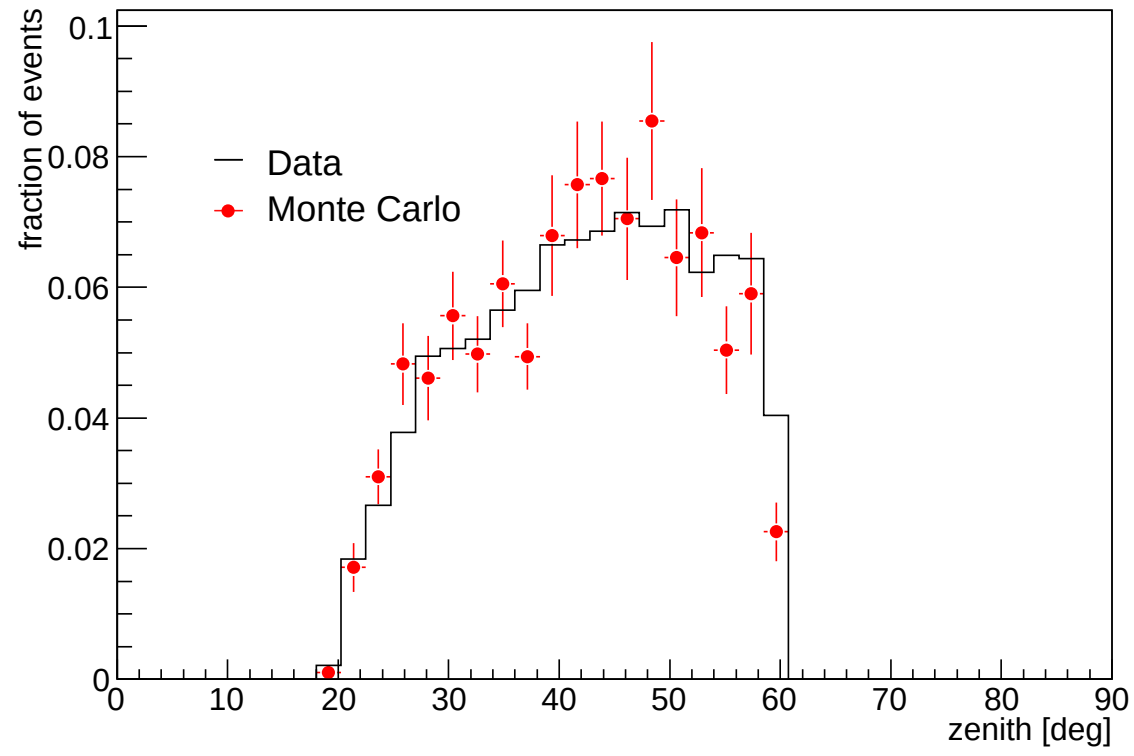

Fig. 4. Data-Monte Carlo Comparison: fraction of hybrid events as a function of the measured zenith angle for the events passing the quality cuts. Both data (solid line) and simulations (solid circles) are shown.

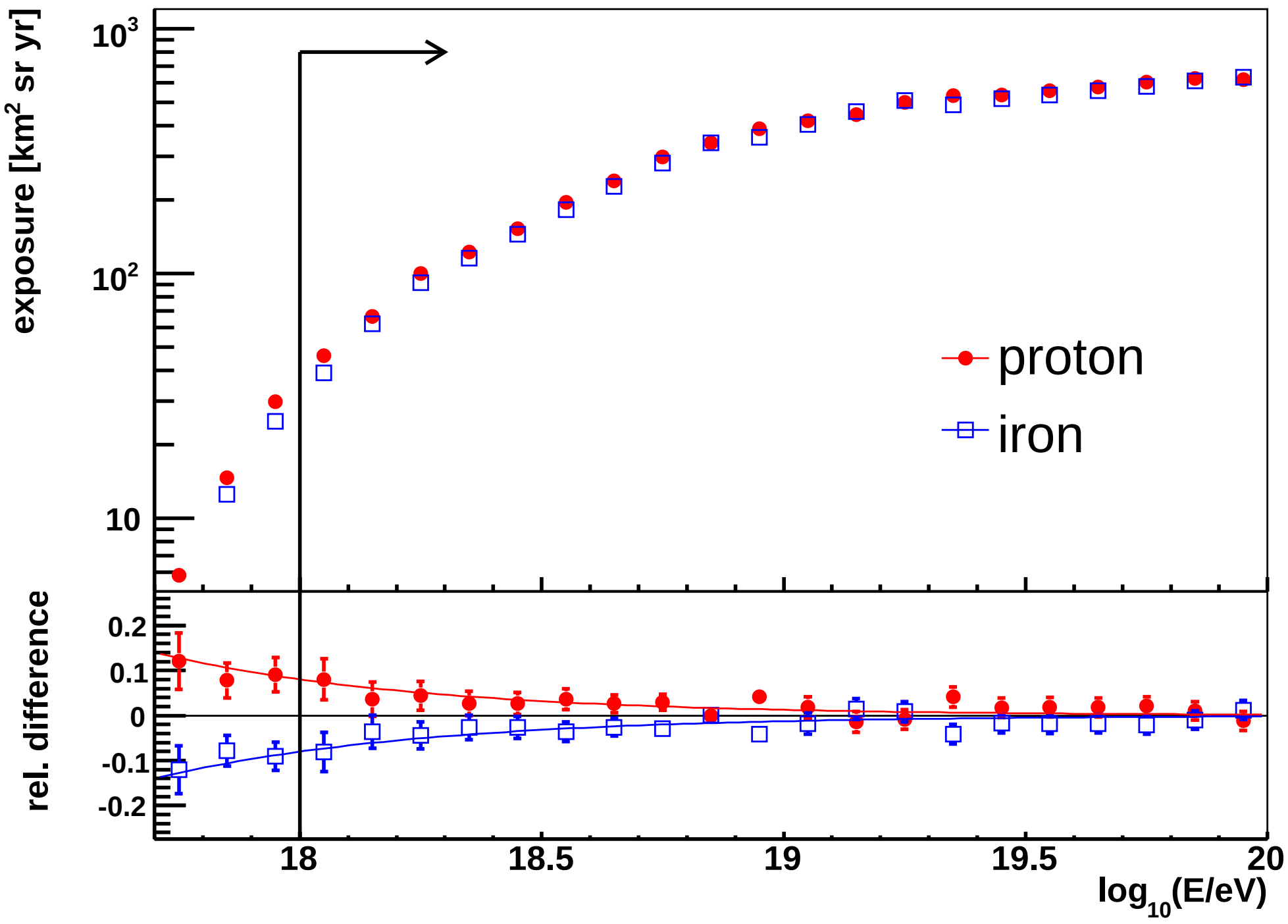

Fig. 5. The hybrid exposure for proton (solid dot) and iron (open squares) primaries derived from Monte Carlo simulation. The relative difference between pure proton(iron) exposure and a mixed composition exposure is shown in the lower panel.

A first set is based on the quality of the geometrical reconstruction:

- reconstructed zenith angle less than 60°;
- station used for the hybrid reconstruction lying within 1500 m from the shower axis;
- energy dependent core-FD site distance according to [17];
- energy dependent field of view according to [18].

A second set of cuts is based on the quality of the reconstructed profile:

- a successful Gaisser-Hillas fit with $\chi^2$/Ndof $< 2.5$ for the reconstructed longitudinal profile.
- minimum observed depth $<$ depth at shower maximum ($X_{max}$) $<$ maximum observed depth;
- events with relative amount of Cherenkov light in the signal less than 50%;
- energy reconstruction uncertainty less than 20%;

A final set of cuts is based on the quality of the atmospheric conditions:

- measurement of atmospheric parameters available [19], [9];
- cloud coverage from Lidar measurements [9] lower than 25%.

The reliability of quality cuts has been checked by comparing the distributions of data and Monte Carlo for all the relevant shower observables. The fraction of selected hybrids events is shown in Figure 3 as a function of time. In this plot both the growing of the hybrid detector and the seasonal trend of the hybrid data taking efficiency are visible. As an example the distributions of the measured zenith angle for both data and Monte Carlo are shown in Figure 4. In this plot only the events passing the quality cuts are shown. Data are in an agreement with simulations.

## V. Results

The hybrid exposure is shown in Figure 5 both for proton (solid dot) and iron (open squares) primaries. The black arrow indicates the region above $10^{18}$ eV where the exposure is used for the measurement of the hybrid spectrum. The exposure has been corrected for a 4% systematic inefficiency derived from the analysis of Central Laser Facility [19] shots. The residual difference between pure proton/iron exposure and a mixed composition (50% proton - 50% iron) one is about 8% at $10^{18}$ eV and decreases down to 1% at higher energies. The dependence of the exposure on the hadronic interaction model has been studied in detail by comparing the exposures obtained respectively with QGSJet-II and Sibyll 2.1 Monte Carlo events. The result shows that this effect is negligible.

The design of the Pierre Auger Observatory with its two complementary air shower detection techniques offers the chance to validate the full Monte Carlo simulation chain and the derived hybrid exposure with air shower observations themselves. Based on this end-to-end comparison the hybrid event rate from data has shown a discrepancy of 8% with respect to Monte Carlo simulations. The exposure has been corrected for half of the observed difference and an upper limit of the systematic uncertainty of about 5% is estimated. Taking into account all these contributions the overall systematic uncertainty on the knowledge of the exposure ranges from 10% at $10^{18}$ eV to 6% above $10^{19}$ eV.

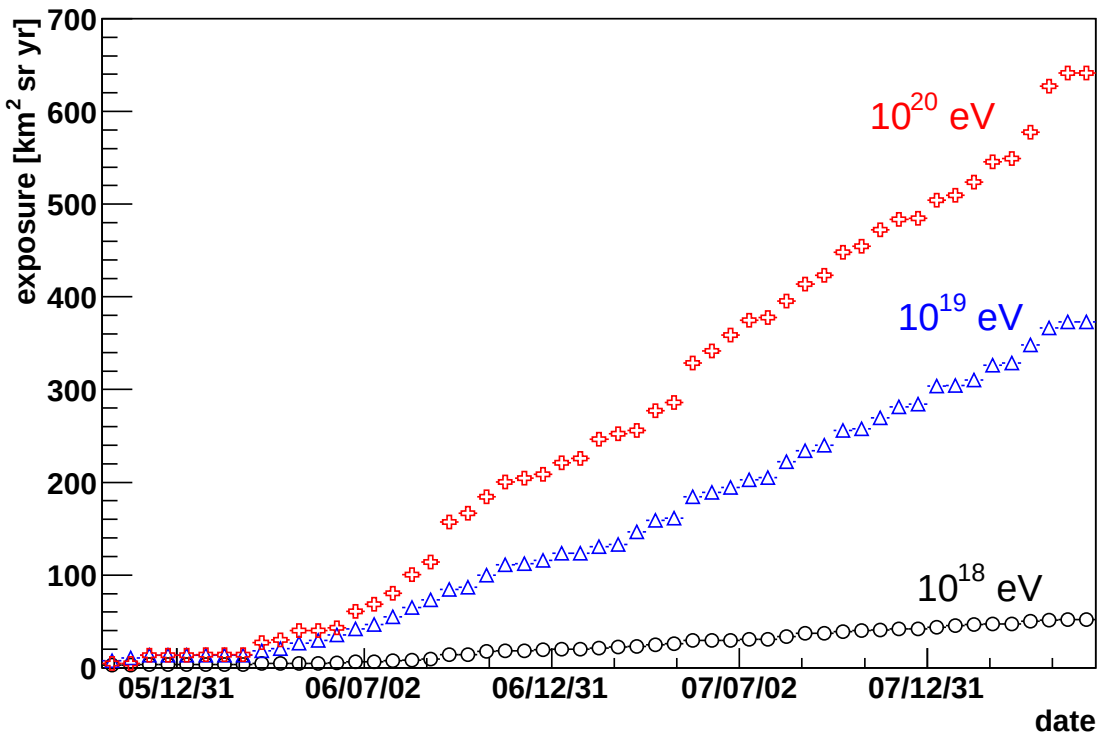

Fig. 6. The growth of the hybrid exposure as a function of time starting from November 2005 up to May 2008 for three different energies.

In Figure 6, the growth of the hybrid exposure as a function of time is shown for three different energies. The increase shown at each energy is not only related to the accumulation of data taking with time. In particular one can easily observe faster changes corresponding to the longer periods in the austral winter. The trend is also affected by the growth of the SD array in the corresponding period. This effect is more marked at higher energies where a bigger hybrid detection volume is accessible with new SD stations.

## VI. Conclusions

A time dependent Monte Carlo simulation has been performed and the exposure of the hybrid detector of the Pierre Auger Observatory has been derived. The use of the monitoring information of the Pierre Auger Observatory allows to follow in detail the changes in the data taking configuration and atmospheric conditions as confirmed by the comparison between data and Monte Carlo. This procedure ensuresa systematic uncertainty on the knowledge of the exposure lower than 10% on the entire energy range used for the measurement of the hybrid spectrum [1].

# Energy scale derived from Fluorescence Telescopes using Cherenkov Light and Shower Universality

**Steffen Mueller*  for the Pierre Auger Collaboration†**

* *Karlsruhe Institute of Technology, Postfach 3640, 76021 Karlsruhe, Germany*
† *Observatorio Pierre Auger, Av. San Martin Norte 304, 5613 Malargüe, Argentina*

***Abstract*. We describe a method to determine the energy scale of the fluorescence detection of airshowers based on the universal shape of longitudinal shower profiles. For this purpose, the ratio of scattered Cherenkov and fluorescence light is adopted as a free parameter while fitting the individual profiles of the longitudinal deposit of the energy to the universal shape. We demonstrate the validity of the method using a Monte Carlo study based on the detector simulation of the Pierre Auger Observatory and estimate systematic uncertainties due to the choice of high energy interaction model and atmospheric conditions.**

## I. Introduction

Knowing the absolute energy scale of cosmic ray detection is important for the interpretation of physics results such as flux, anisotropy, or composition. At the Pierre Auger Observatory, the energy measured with the fluorescence detector is used to calibrate that of the surface detector [1]. Previous experiments that consisted of a surface array used Monte Carlo simulations for their energy calibration.

In air shower detection with fluorescence telescopes, the atmosphere acts as a calorimeter. The amount of emitted fluorescence light is proportional to the energy deposit in the atmosphere. The light yield is measured in laboratory experiments with a precision that is at present typically $15\%$ [2].

Here, we describe a method to obtain the overall normalization of the fluorescence yield directly from air shower measurements. This method makes use of the universality of the shape of the longitudinal shower profiles of the energy deposit in the atmosphere. It is also dependent on our ability to reliably calculate the Cherenkov light contribution (given the electron number and energy spectra). Only the *relative* fluorescence spectrum is needed, which is known with good precision from laboratory experiments.

As an air shower develops in the atmosphere, a beam of Cherenkov light builds up along the axis of the shower and undergoes Rayleigh and aerosol scattering. In general the scattered Cherenkov light which is observed from a certain point in the shower will have been originally emitted at an earlier stage of shower development. The result is a very different longitudinal light profile from that of the isotropically emitted fluorescence light. Therefore, the shape of the reconstructed longitudinal profile of the energy deposit depends on the assumed composition of the different contributions to the measured light. We modify the fluorescence light yield in the reconstruction of the longitudinal profile to change the light composition in such a way that the energy deposit profile matches the profile expected from universality.

## II. Universality of Air Shower Profiles

The energy spectra of shower electrons and the differential energy deposit have been shown to be universal as a function of shower age, $s = 3X/(X + 2X_{\max})$ [3]–[9]. As a result, the shape of energy deposit profiles have been studied for universality when plotted as a function of age. It was found that the profile shape varied much less when plotted in terms of the depth relative to shower maximum, $\Delta X = X - X_{\max}$. Figure 1 shows many normalized energy deposit profiles in $\Delta X$ that were simulated with proton primaries using three different high-energy interaction models at $10^{19}$ eV. In $\Delta X$, the majority of normalized profiles fall within a narrow band.

Consider the average of normalized energy deposit profiles $U_i(\Delta X)$ for a single interaction model and primary particle. Then figure 2 shows the absolute deviations $\delta_i(\Delta X)$ of each average profile from the mean $\langle U(\Delta X)\rangle$ of the average profiles

$$\begin{aligned} U_i(\Delta X) &= \left\langle \left(\frac{dE}{dX}\right) \Big/ \left(\frac{dE}{dX}\right)_{\max} \right\rangle (\Delta X) \\ \delta_i(\Delta X) &= U_i(\Delta X) - \langle U(\Delta X)\rangle \,. \end{aligned}$$

Nowhere does the total systematic difference rise above $3\%$ from the mean and it stays below one percent after the shower maximum. The equivalent plot for shower age shows deviations of up to $5\%$ from the mean both before and after the shower maximum. Due to the weak dependence on primary composition, interaction model and primary particle energy, the average profile $U(\Delta X)$ is henceforth referred to as the Universal Shower Profile (USP). The measurement of the energy scale of fluorescence detection with the method described below is most susceptible to systematic differences in the tail of the USPs for different parameters (cf. figure 3c).

There is a slight dependence of the shape of the energy deposit profile on the primary energy. This effect is relevant for this work only within the uncertainty of the reconstructed primary energy because the USP was recalculated for each event from simulations at the estimated energy. The dependence of the shape on the primary energy introduces a negligible systematic uncertainty (cf. table I).

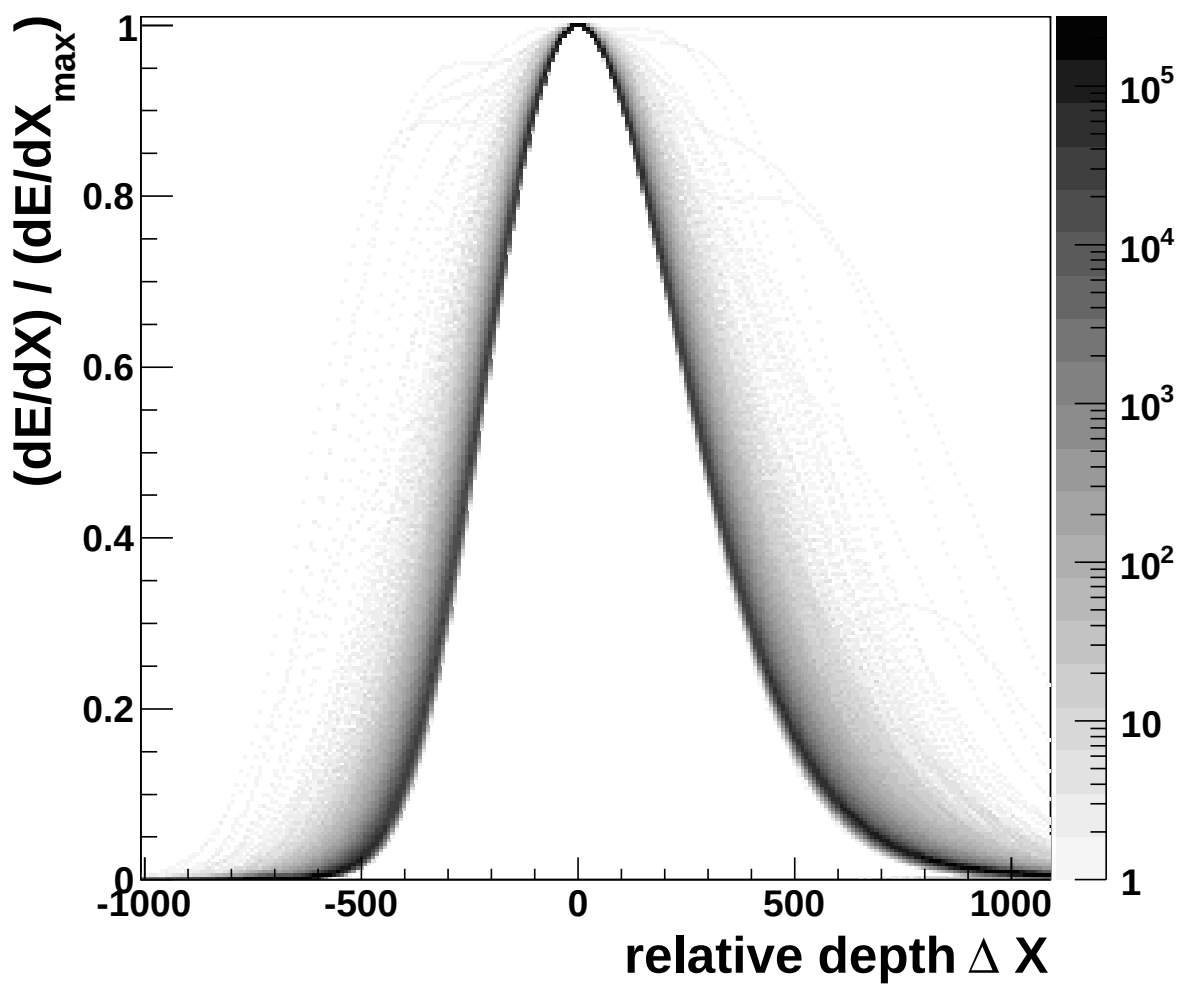

Fig. 1. Superposition of 30000 energy deposit profiles in $\Delta X$. The CONEX [10] simulations are for proton primaries. Equal numbers of showers were generated with the QGSJet, QGSJetII.03, and Sibyll interaction models [11]–[13].

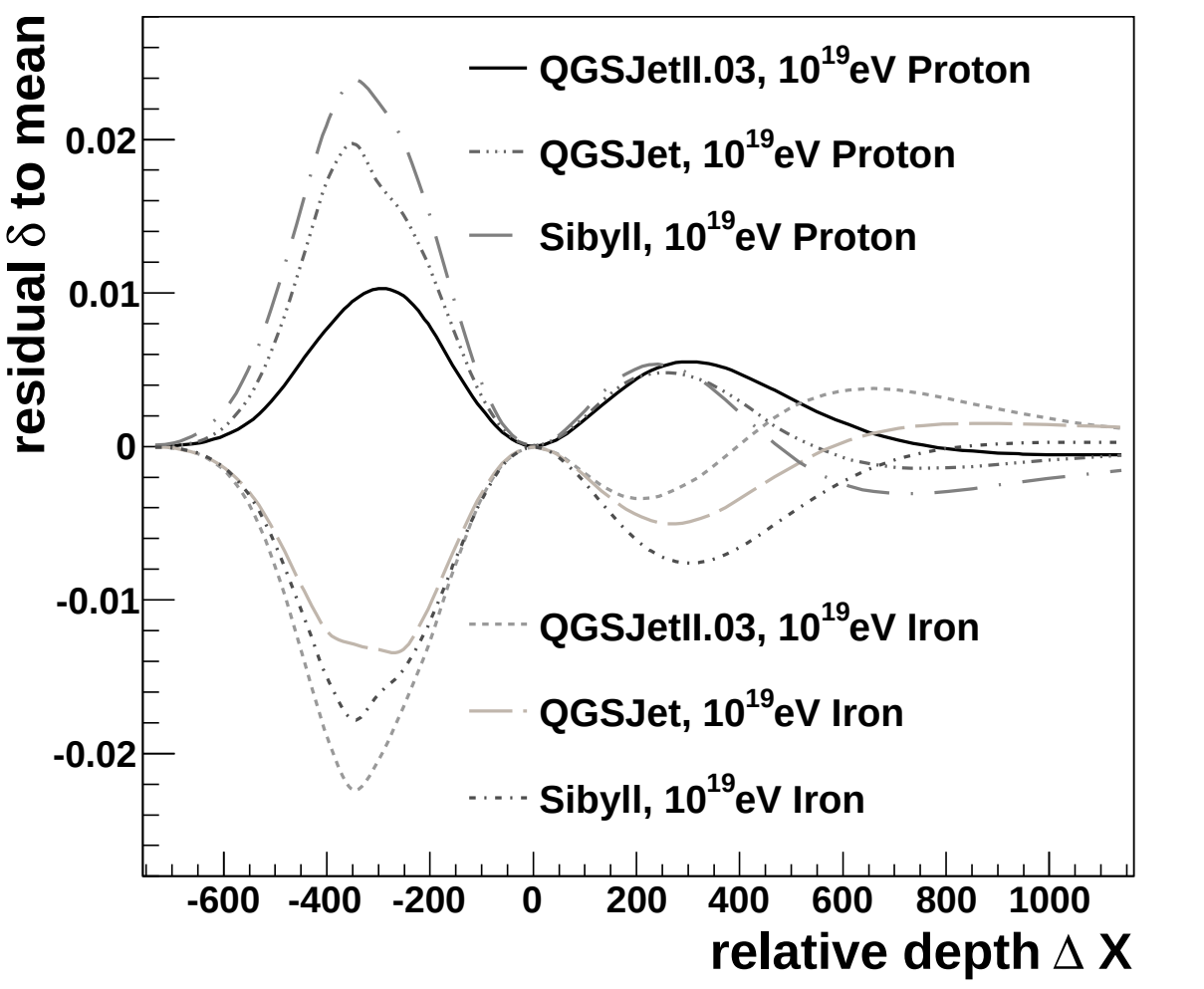

Fig. 2. Residuals $\delta(\Delta X)$ of universal shower profiles for various interaction models and primaries to the mean of the profiles.

## III. Method

The longitudinal profile reconstruction [14] of the Offline software framework [15] was extended with an additional free parameter $f$ so that the fluorescence yield becomes $Y^f = Y^f_{\rm lab}/f$ where $Y^f_{\rm lab}$ is the fluorescence yield currently used in the standard shower reconstruction by the Auger Collaboration [1]. This fluorescence yield is a parameterization of laboratory measurements, including the corresponding pressure dependence. Since $Y^f$ is inversely related to $f$, a change in $f$ corresponds to a proportional change in the reconstructed shower energy.

A set of showers is reconstructed many times while varying $f$. A low $f$ corresponds to assuming a large fluorescence light yield and implies that fewer electrons are required in the shower to produce the observed fluorescence light. Since a smaller number of particles emits less Cherenkov light, the fraction of the measured light that is reconstructed as Cherenkov light is reduced accordingly.

The majority of detected fluorescence photons has not been scattered in the atmosphere before reaching the detector. Therefore, the point on the shower axis from which fluorescence light is observed is also the point at which it was emitted. Showers with significant contributions of direct Cherenkov light are not selected for this analysis (see below). Thus, the bulk of the observed Cherenkov light has propagated along the shower axis before being scattered towards the detector on molecules or aerosols. This means that the detected Cherenkov light carries information from a different stage of shower development than the fluorescence light observed from the same direction. This gives us a handle to change the shape of the reconstructed longitudinal profile of the energy deposit for a given observed light profile by modifying the fluorescence yield scale factor $f$.

The effect of a modified $f$ parameter on the reconstructed light composition is demonstrated with an example in figure 3a and 3b. The measured light profile is unchanged. But with higher fluorescence yield in 3a, the contribution of Cherenkov light is suppressed. Conversely, it is increased due to the reduced fluorescence yield in 3b.

At the same time, a modified $f$ changes the shape of the reconstructed energy deposit profile as shown in figure 3c. Since the shape is known from universality considerations, a $\chi^2$ minimization can be used to fit each profile to the universal shape in dependence of $f$.

Each event is assigned an uncertainty that is a combination of the uncertainty from the $\chi^2$ minimization and several propagated uncertainties. These include the uncertainties on the direction of the shower axis, the spread of the showers that make up the Universal Shower Profile, and the uncertainty on the aerosol attenuation lengths. The fit is repeated twice for each of these parameters: once after increasing and once after decreasing each parameter by one standard deviation. The resulting difference to the default result is the propagated uncertainty.

## IV. Results

To test the method, a set of showers that roughly corresponds to five years of Auger data was simulated with energies between $10^{18}$ and $10^{20}$eV. The simulation setup follows that used for the Auger fluorescence detector exposure calculation [16]. Basic quality cuts such as requiring an energy resolution better than 20% and an $X_{\max}$ resolution better than $40\text{g}/\text{cm}^2$ were applied. Additionally, since the forward peaked nature of direct Cherenkov light introduces a strong susceptibility to the uncertainties of geometry reconstruction, showers with a significant contribution of direct Cherenkov light were not used for the analysis. This was implemented by selecting showers with a minimum viewing angle in excess of 20°. The minimum viewing angle is the minimum angle between the shower axis and any vector between a point in the observed profile and the fluorescence detector.

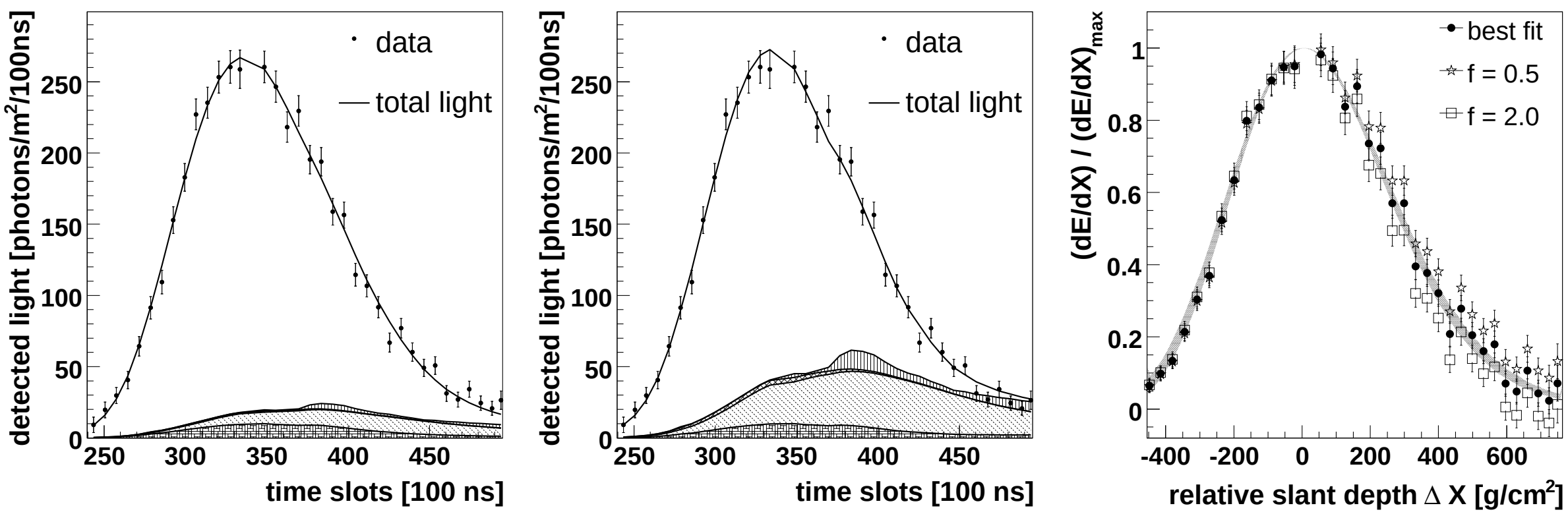

(a) $f = 0.5$ (b) $f = 2.0$ (c)

Fig. 3. Example event (Los Morados detector, run 1392, event 2886). (a)/(b) Measured light profile with reconstructed light components for two modified yield scale factors. Fluorescence light , Cherenkov light (direct , Mie scattered , Rayleigh scattered ), multiply scattered light ; (c) Normalized, reconstructed energy deposit profiles. Grey band: Universal shower profile with uncertainty band. Graphs: energy deposit profiles for different values of the yield scale factor

Conversely, the showers were required to have significant contributions of Rayleigh scattered Cherenkov light, and a long profile that includes both regions in slant depth where fluorescence light and regions where Cherenkov light dominate the measured light flux.

This requirement was implemented as a two-dimensional cut on the profile length after the shower maximum and a quantity $R$

$$R = \rho(X_{\text{up}}) \cdot \left(1 + \cos^2 \psi\right) .$$

It is the product of atmospheric density $\rho$ in the deepest visible part of the shower track $X_{\text{up}}$ and the angular dependence of Rayleigh scattering given the viewing angle $\psi$. Thus $R$ is a measure for the amount of Cherenkov light scattered from the end of the profile towards the telescope.

The Monte Carlo simulation was carried out for three different fluorescence yields:

- The laboratory measurement $Y^f_{\text{default}} = Y^f_{\text{lab}}$ (corresponding to $f_{\text{true}} = 1.0$),
- an increased fluorescence yield $Y^f_{\text{high}} = Y^f_{\text{lab}}/0.8$ ($f_{\text{true}} = 0.8$),
- and a lowered fluorescence yield $Y^f_{\text{low}} = Y^f_{\text{lab}}/1.2$ ($f_{\text{true}} = 1.2$).

In the shower reconstruction, the fluorescence yield was $Y^f_{\text{lab}}/\tilde{f}$ with the fit parameter $\tilde{f}$.

For the selected set of simulated showers, the resulting, reconstructed fluorescence yield scale factors $\tilde{f}$ are weighted with their respective uncertainties. The distribution of these weighted scale factors is shown in figure 4 for three different input values of $f_{\text{true}}$. As can be seen, we are able to recover the true yield with good accuracy. This shows that the method is sensitive to a true fluorescence yield which differs from the assumed yield $Y^f_{\text{lab}}$ because the reconstructed scale factor $\tilde{f}$ has no bias relating to the input parameter $f_{\text{true}}$. The width of the distributions, however, shows that a large number of suitable showers is required for the analysis.

The systematic uncertainties (table I) from various sources were taken into account by repeating the full procedure with various input parameters modified by their respective systematic uncertainties. For the systematics of the method, aerosols play a particularly important role. Both aerosol attenuation and scattering of Cherenkov light on aerosols are non-trivial effects that change the shape of the reconstructed energy deposit profile. The largest contribution is due to the uncertainties of the vertical aerosol optical depth (VAOD) profile. Since the available uncertainty bounds include both statistical and systematic effects, we estimate an upper limit for the systematics on $f$ of about $\pm 7\%$. Another significant systematic uncertainty is introduced by the parameters of the aerosol phase function (APF) which describes the angular dependence of scattering on aerosols [17]. Its parameter $g$ is a measure for the asymmetry of scattering, whereas the APF parameter $f$ determines the relative strength of forward and backwards scattering. The contribution from the exponent $\gamma$ describing the wavelength dependence of light attenuation due to aerosols is small. Likewise, the slight energy dependence of the shape of the universal shower profile leads to an uncertainty of less than one percent. Using various models or compositions for calculating the universal shower profile yields another contribution to the total of $\pm 1\%$ and $\pm 3\%$ respectively. Two different

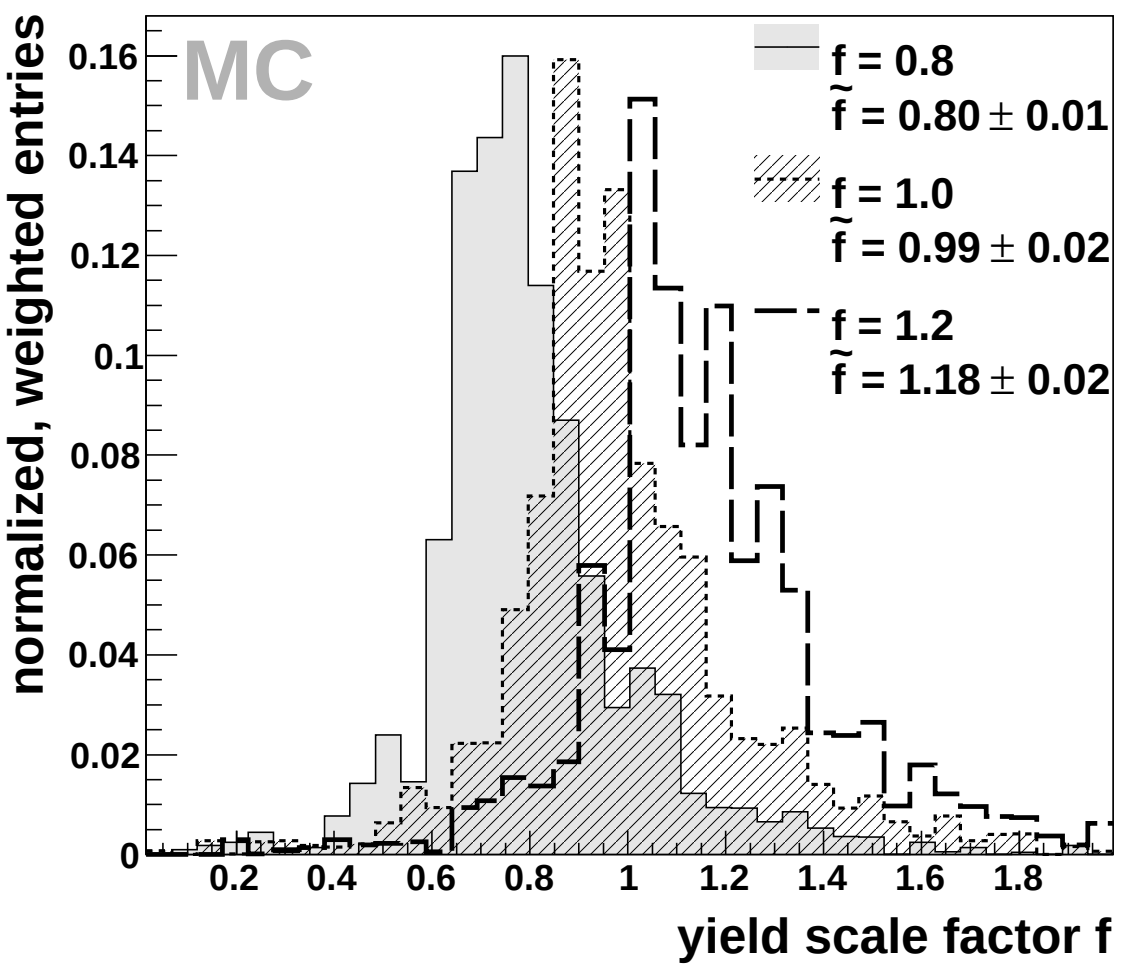

Fig. 4. Reconstructed, weighted yield scale factor distributions for three different input values of the scale factor $f$

TABLE I

UNCERTAINTIES OF THE SCALE FACTOR (SEE TEXT)

| Source | Uncertainty [%] |
|---|---|
| APF: $g$ | $+5, -3$ |
| APF: $f$ | $\pm 0.4$ |
| wavelength dependence $\gamma$ | $+0.0, -0.2$ |
| VAODs | $\approx \pm 7$ |
| multiple scattering | $\pm 0.5$ |
| energy reconstruction | $+0.4, -0.5$ |
| USP had. int. model | $\pm 1$ |
| USP composition | $\pm 3$ |

parameterizations for the multiple scattering of light in the atmosphere [18], [19] produce yield scale factors that differ by 1%. If added in quadrature, these effects add up to a total expected systematic uncertainty of $8-9\%$.

## V. CONCLUSIONS

We introduced a new method of measuring the energy scale of fluorescence detectors using the universality of shape of the longitudinal shower profile. Its applicability and sensitivity was demonstrated using Monte Carlo simulations of air showers and the detector of the Pierre Auger Observatory. The measurement of the energy scale uses air shower data to determine the absolute fluorescence yield scale directly, and only requires a laboratory measurement of the relative fluorescence spectrum.

The simulated fluorescence yields were reproduced to very good accuracy. The systematic uncertainties of this method could potentially allow for a fluorescence yield determination with a precision better than 10%. The application of this method to Auger data is in progress.

# Acknowledgements

The successful installation and commissioning of the Pierre Auger Observatory would not have been possible without the strong commitment and effort from the technical and administrative staff in Malargüe.

We are very grateful to the following agencies and organizations for financial support:

Comisión Nacional de Energía Atómica, Fundación Antorchas, Gobierno De La Provincia de Mendoza, Municipalidad de Malargüe, NDM Holdings and Valle Las Leñas, in gratitude for their continuing cooperation over land access, Argentina; the Australian Research Council; Conselho Nacional de Desenvolvimento Científico e Tecnológico (CNPq), Financiadora de Estudos e Projetos (FINEP), Fundação de Amparo à Pesquisa do Estado de Rio de Janeiro (FAPERJ), Fundação de Amparo à Pesquisa do Estado de São Paulo (FAPESP), Ministério de Ciência e Tecnologia (MCT), Brazil; AVCR AV0Z10100502 and AV0Z10100522, GAAV KJB300100801 and KJB100100904, MSMT-CR LA08016, LC527, 1M06002, and MSM0021620859, Czech Republic; Centre de Calcul IN2P3/CNRS, Centre National de la Recherche Scientifique (CNRS), Conseil Régional Ile-de-France, Département Physique Nucléaire et Corpusculaire (PNC-IN2P3/CNRS), Département Sciences de l'Univers (SDU-INSU/CNRS), France; Bundesministerium für Bildung und Forschung (BMBF), Deutsche Forschungsgemeinschaft (DFG), Finanzministerium Baden-Württemberg, Helmholtz-Gemeinschaft Deutscher Forschungszentren (HGF), Ministerium für Wissenschaft und Forschung, Nordrhein-Westfalen, Ministerium für Wissenschaft, Forschung und Kunst, Baden-Württemberg, Germany; Istituto Nazionale di Fisica Nucleare (INFN), Ministero dell'Istruzione, dell'Università e della Ricerca (MIUR), Italy; Consejo Nacional de Ciencia y Tecnología (CONACYT), Mexico; Ministerie van Onderwijs, Cultuur en Wetenschap, Nederlandse Organisatie voor Wetenschappelijk Onderzoek (NWO), Stichting voor Fundamenteel Onderzoek der Materie (FOM), Netherlands; Ministry of Science and Higher Education, Grant Nos. 1 P03 D 014 30, N202 090 31/0623, and PAP/218/2006, Poland; Fundação para a Ciência e a Tecnologia, Portugal; Ministry for Higher Education, Science, and Technology, Slovenian Research Agency, Slovenia; Comunidad de Madrid, Consejería de Educación de la Comunidad de Castilla La Mancha, FEDER funds, Ministerio de Ciencia e Innovación, Xunta de Galicia, Spain; Science and Technology Facilities Council, United Kingdom; Department of Energy, Contract No. DE-AC02-07CH11359, National Science Foundation, Grant No. 0450696, The Grainger Foundation USA; ALFA-EC / HELEN, European Union 6th Framework Program, Grant No. MEIF-CT-2005-025057, European Union 7th Framework Program, Grant No. PIEF-GA-2008-220240, and UNESCO.